\documentclass[conference]{IEEEtran}
\IEEEoverridecommandlockouts
\usepackage{cite}
\usepackage[top=2cm, bottom=2cm, left=2cm, right=2cm]{geometry}
\usepackage{amsmath,amssymb,amsfonts}
\usepackage{graphicx}
\usepackage{textcomp}
\usepackage{xcolor}
\usepackage{url}
\usepackage{amsmath}
\usepackage{algorithm}
\usepackage{algorithmicx}
\usepackage{algpseudocode}
\usepackage{booktabs}
\usepackage{subfigure}
\usepackage{verbatim}
\usepackage[]{footmisc}
\usepackage{flushend}
\usepackage{dashrule}
\usepackage[OT1]{fontenc}
\usepackage[colorlinks,
linkcolor=blue,
anchorcolor=blue,
citecolor=red
]{hyperref}
\def\BibTeX{{\rm B\kern-.05em{\sc i\kern-.025em b}\kern-.08em
		T\kern-.1667em\lower.7ex\hbox{E}\kern-.125emX}}
\begin{document}
	
	\floatname{algorithm}{Algorithm}
	\renewcommand{\algorithmicrequire}{\textbf{Input:}}
	\renewcommand{\algorithmicensure}{\textbf{Output:}}
	
	\title{SpotTune: Leveraging Transient Resources for Cost-efficient Hyper-parameter Tuning in the Public Cloud}
	
	\author{
		\IEEEauthorblockN{Yan Li, Bo An, Junming Ma, Donggang Cao*\thanks{*Corresponding author.}, Yasha Wang, Hong Mei}
		\IEEEauthorblockA{
			\textit{Key Lab of High-Confidence Software Technology (Peking University), Ministry of Education, Beijing, China} \\
			\{yan.l, anbo, mjm520, caodg, wangyasha, meih\}@pku.edu.cn
		}
	}
	
	\maketitle
	
	\begin{abstract}
	Hyper-parameter tuning (HPT) is crucial for many machine learning (ML) algorithms. But due to the large searching space, HPT is usually time-consuming and resource-intensive. Nowadays, many researchers use public cloud resources to train machine learning models, convenient yet expensive. How to speed up the HPT process while at the same time reduce cost is very important for cloud ML users. In this paper, we propose SpotTune, an approach that exploits transient revocable resources in the public cloud with some tailored strategies to do HPT in a parallel and cost-efficient manner. Orchestrating the HPT process upon transient servers, SpotTune uses two main techniques, fine-grained cost-aware resource provisioning, and ML training trend predicting, to reduce the monetary cost and runtime of HPT processes. Our evaluations show that SpotTune can reduce the cost by up to 90\% and achieve a 16.61x performance-cost rate improvement.
	\end{abstract}
	
	
	\section{Introduction}\label{sec_intro}
	Machine learning (ML) has become an important workload for both research and industry that tries to learn insight from the collected dataset. An ML model usually consists of a set of parameters, which can be updated during the training process. Yet there are also some non-trainable parameters that need to be determined by developers before training, e.g., number of hidden layers, learning rate, degree of regularization, kernel function of SVM, etc., namely hyper-parameters. In order to predict outcomes for new-coming data items more precisely, developers need to repeat the training processes with different hyper-parameter (HP) settings until getting the optimal model. This trial-and-error procedure is called hyper-parameter tuning (HPT).
	
	Tuning hyper-parameters for an ML model is resource-intensive and expensive, requiring trying models with different configurations. The emerging public cloud has unlimited computation capacity, facilitating many developers to do HPT in parallel by renting a set of machines, which is convenient and fast, but very expensive. Many cloud vendors offer a great opportunity: transient availability of cheap but revocable resources \cite{harlap2018tributary,harlap2017proteus}. For example, AWS spot instances\cite{awsspot} and Google Cloud's preemptible resources\cite{googlecloud} often provide users with machines at a 70\%-80\% discount compared with the regular price of the on-demand instances, but with the risk that some of them might be taken away at any time.
	
	Intuitively, using these dynamic resources to do HPT in parallel, i.e., launching a bunch of transient servers simultaneously, each of which tries a different hyper-parameter setting, could be both fast and cheap. But to use it robustly and efficiently, there are three key issues to consider.
	
	\textbf{Handle the interruption properly.} Dynamic resources are cheap for a reason: they could be revoked at any time by the cloud vendors, which then interrupts the ML training. If no fault-tolerant strategies are applied, the intermediate training results would lose when encountering an interruption.
	
	\textbf{Cost-aware resources provisioning.} There are hundreds of instance types in the public cloud dynamic infrastructure market, each of which has a different price and computation ability. Besides, the price of transient resources is variable. Facing such complex cloud markets, the user could be confused about how to provision resources to reduce the overall cost.
	
	\textbf{Make full use of the elasticity of the public cloud.} The public cloud provides the ability of elasticity, i.e., acquiring/releasing resources on-demand and only paying for the used computation capacity. HPT is well-suited for this mechanism because, after the exhaustive searching of the hyper-parameters, only a small part of the models will be left. If the models initiated with bad hyper-parameters could be identified early, computation capacity could be released in time to lower the cost.
	
	To tackle the above issues, this paper describes SpotTune, an orchestrating system that combines a fine-grained cost-aware resource provisioning strategy with an ML training trend predicting method to leverage transient cloud resources for HPT. SpotTune checkpoints the intermediate training data to the remote persistent storage when receiving the pre-delivered revocation notice, thus mitigating the risk of interruption, and its cost-efficient effect is powered by the following two principal techniques:
	
	\textbf{Fine-grained Cost-aware Resource Provisioning.} SpotTune provisions resources for HPT jobs by computing the expected cost at a fine granularity. Via predicting spot instance revocation probability and online profiling HPT job performance, SpotTune computes the expected cost of a single iteration in the next hour and selects the instance that incurs the least cost to deploy HPT jobs.
	
	\textbf{Early-Shutdown based on Training Trend Predicting.} SpotTune models the training process as a predictable curve, predicting the training trend of a model using the metrics (e.g., validation loss) collected during the training. Generally speaking, it can judge whether a model is promising or not using partial metrics thus could release resources running bad models in time. It just lets the models initiated with bad hyper-parameters stop before they take more resources to do the unnecessary computation, denoted as early-shutdown in this paper.
	
	This paper makes the following four primary contributions.
	\begin{enumerate}
		\item It describes SpotTune, an orchestrating system that takes advantage of transient cloud resources to do HPT.
		\item It introduces a novel transient resources provisioning approach by computing the expected cost at a fine granularity.
		\item It proposes an enhanced spot instance revocation probability predicting method.
		\item It introduces a new ML model training trend predicting method, modeling the training process as a staged predictable curve.
	\end{enumerate}
	
	The reminder of this paper is organized as follows. Section \ref{section_2} presents the background and motivation of this paper. Section \ref{section_3} introduces the design of SpotTune. Section \ref{section_5} evaluates SpotTune. Section \ref{section_6} narrates some discussions of our system. Section \ref{sec_related_work} briefly introduces the related work.  Section \ref{section_7} concludes this paper.
	
	\section{Background and Motivation}\label{section_2}
	
	This section briefly introduces the background, including existing transient cloud resource markets and their principal traits, hyper-parameter tuning, and the motivation of combining them together.
	
	\subsection{Transient Cloud Resources}\label{subsec_resource}
	
	Resources provisioned for business-critical workloads could be available at a discount in public pay-as-you-go clouds, creating an opportunity for other workloads. But, those resources may be taken back when needed by cloud vendors again. Taking Amazon AWS EC2 Spot Instance as an example, this subsection describes transient cloud resources and the opportunity they provide us.
	
	Amazon Elastic Compute Cloud (Amazon EC2)\cite{awsec2} is a web service that provides secure, resizable compute capacity in the cloud. EC2 offers \textit{on-demand instances}, which are reliable VMs charged at a per-second rate. Besides the reliable resources, EC2 also provides cheaper (usually 70\%-80\% lower than the on-demand resources) yet preemptible resources as \textit{spot instances}. Spot instances could be revoked at any time, thus providing users a trade-off between reliability and cost savings.
	
\begin{figure}[!htbp]
		\centerline{\includegraphics[width=.5\textwidth]{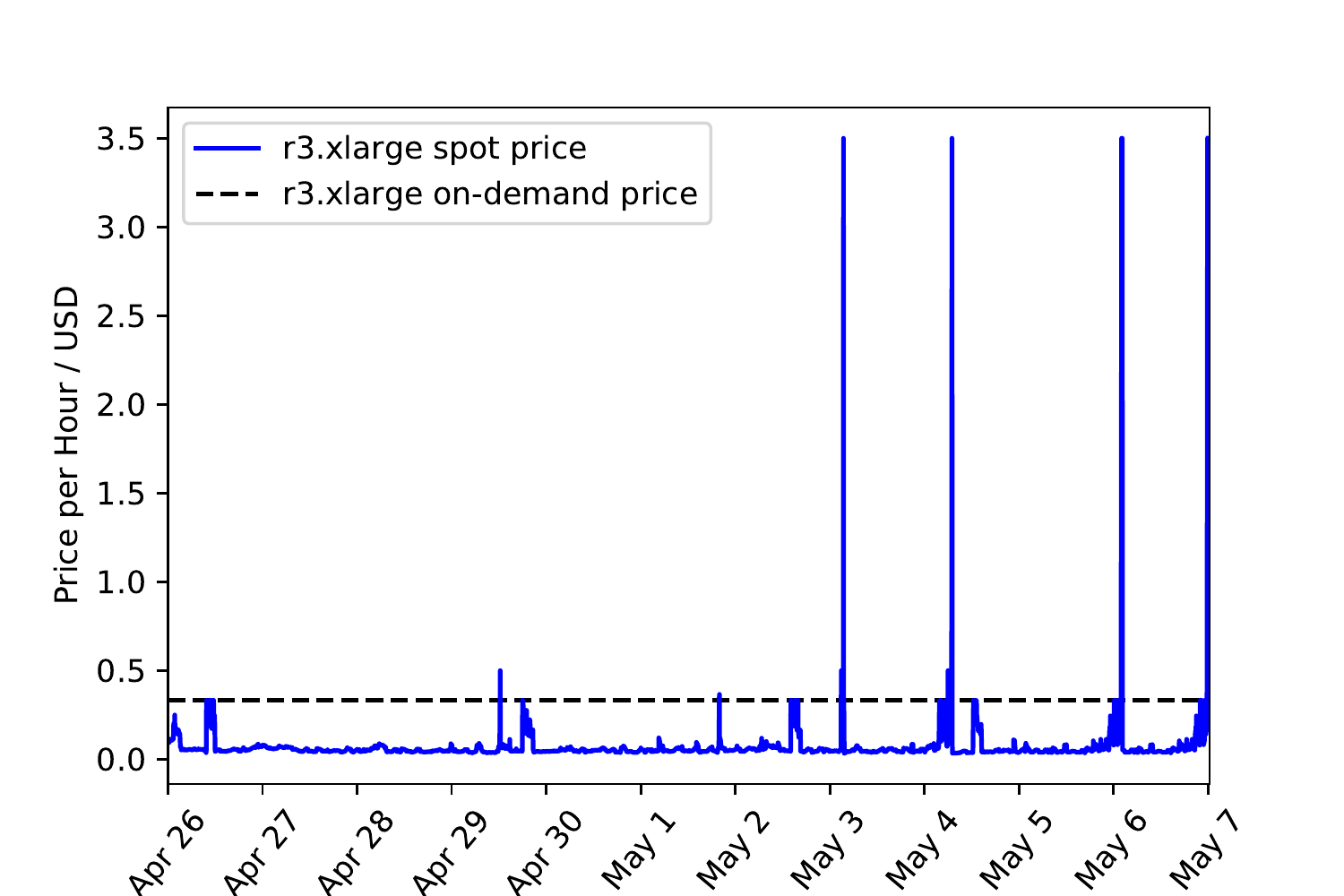}}
		\caption{AWS Spot Prices. Spot prices of instance type \textit{r3.xlarge} across eleven days. The unchanging on-demand price of \textit{r3.xlarge} is shown as well.}
		\label{fig_r3.xlarge}
	\end{figure}

	There are several characteristics of the AWS EC2 spot instance that may affect the monetary cost and the possibility of revocation. First, different instance types have different spot markets\cite{harlap2018tributary,harlap2017proteus}. Second, price fluctuations among different markets are barely correlated. Third, users could specify a \textit{maximum price} when acquiring a spot instance. It could be considered as the maximum price the user would like to pay for it. Once the spot market price is over the user's \textit{maximum price}, the instance would be revoked by AWS. Fourth, the user is charged at a per-second rate with the spot market price (not the maximum price) with an exception: users can get a full refund if the acquired instance is revoked in its first instance hour\cite{ec2spotbonus}. Besides, AWS provides a service that delivers termination notices\cite{awsspotrevoke}, which are issued two minutes before the interruption. Otherwise, there is no refund. In fact, AWS used the term \textit{bid} before 2017 and replaced it with \textit{maximum price} after. But the refund mechanism still exists. Figure \ref{fig_r3.xlarge} shows an example of spot price changes over eleven days.
	
	Collectively, there are two kinds of strategies to use EC2 spot instances. The first kind of strategy bid\footnote{Despite there is no longer the concept of \textit{bid} any more now, the charging and refund mechanism still exists. And as these works are based on the concept of \textit{bid}, we still use the term \textit{bid} here.} \textbf{carefully} to minimize or mitigate the risks of being preempted. Zheng et al.\cite{zheng2015bid} synthetically study the bidding strategies, deriving optimal bidding strategies for different job requirements. Sharma et al.\cite{sharma2017portfolio} present the notion of server portfolios that is based on financial portfolio modeling for reducing the effect of preemption. Subramanya et al.\cite{subramanya2015spoton} present SpotOn, automatically selecting a spot market and fault-tolerance mechanism to mitigate the impact of spot revocations. The other strategies usually just a little higher than the current spot market price, to maximize the odds of being revoked within the first hour, thus enjoying the refund bonus. This kind of bidding strategy is named as \textbf{aggressive} bidding. Harlap et al.\cite{harlap2018tributary} present Tributary, \textit{borrowing} the to-be-preempted machines to satisfy a given service SLO with a lower cost.
	
	\subsection{Hyper-parameter Tuning of ML Models}\label{subsec_tune}
\begin{figure}[!t]
		\centerline{\includegraphics[width=.5\textwidth]{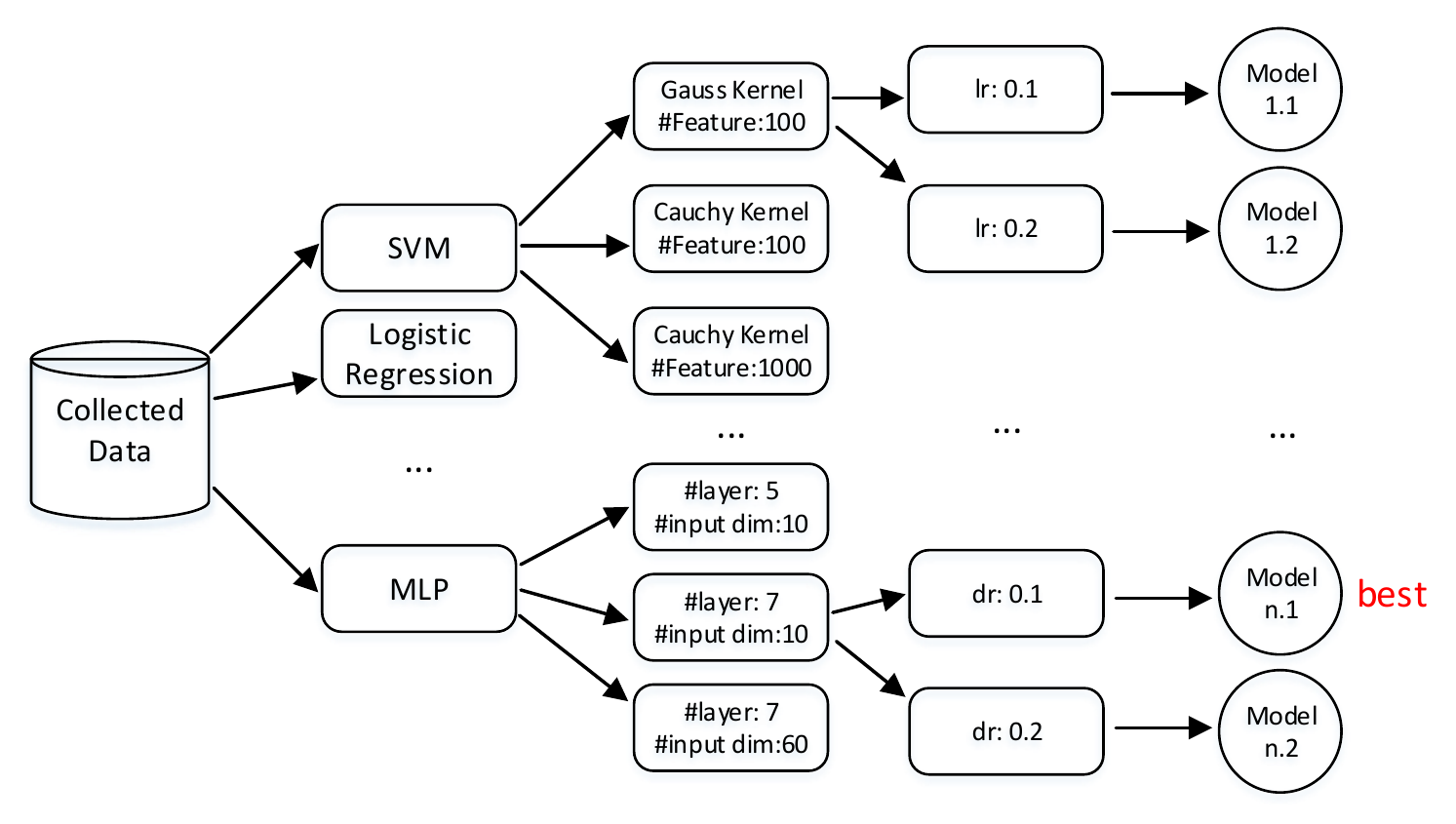}}
		\caption{An Example of \textbf{HPT}. After an exhaustive searching, only one optimal model is left. Here `lr' denotes a learning rate parameter, `dr' a dropout rate parameter, `\#layer' the number of layers in an MLP model, `\#input dim' the dimension of the input data, and `\#Feature' the number of features to use.}
		\label{fig_hpt}
	\end{figure}

	A machine learning model basically consists of a set of trainable parameters, which could be updated during the training process until the pre-defined model fits the training data. There are also some non-trainable parameters that should be determined by developers manually before training, namely hyper-parameters. HPT is, in fact, a process of selecting the best model among the HP space based on domain knowledge and trial-and-error. Figure \ref{fig_hpt} shows an example of HPT.
	
	HPT can be done using various searching algorithms\cite{luo2016review,lu2019hyper,shahriari2015taking,snoek2012practical,bergstra2012random,mantovani2015effectiveness,sparks2015automating} over the space of the hyper-parameters, which means that the user launches multiple ML jobs, each performing full training using one hyper-parameter setting. Because the configuration space to search is usually large, this process is rather computationally expensive, especially in the context of public pay-as-you-go clouds.
	
	Recent works\cite{peng2018optimus,zhang2017slaq} have proved the predictability of ML model training, modeling the training process as a predictable curve. This mechanism could help lower the cost of HPT if combined with the elasticity and pay-as-you-go nature of the public cloud. That is, if bad models could be identified early by predicting the model quality with partially collected metrics (e.g., validation loss), the computation capacity could be released in time.
	
	SpotTune follows this idea, modeling the ML training process as a staged predictable curve to early release machines running bad models.
	
	\subsection{Tuning with Transient Resources}
	
	Inspired by the properties of transient cloud resources, and the nature of HPT jobs, we present SpotTune, addressing the issues of the transient resources opportunity, the cloud elasticity and the ML training predictability to orchestrate HPT upon transient cloud resources in a parallel and cost-effective manner. Section \ref{section_3} describes the design of SpotTune. To our knowledge, SpotTune is the first system leveraging transient resources in the public cloud to do HPT.
	
	\section{SpotTune Design}\label{section_3}
	
	\begin{figure}[!t]
		\centerline{\includegraphics[width=.5\textwidth]{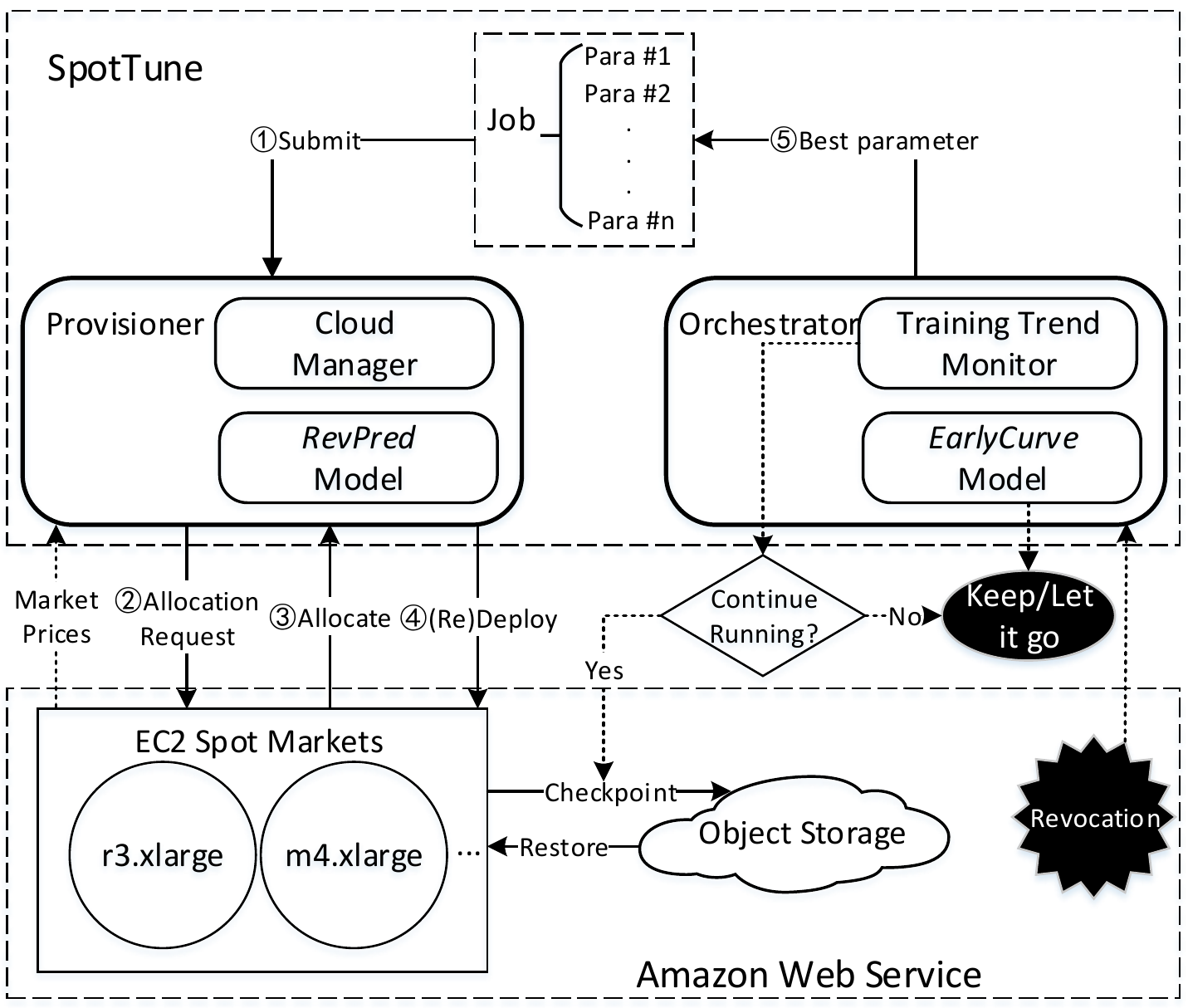}}
		\caption{SpotTune Overview. SpotTune consists of two important components, the Provisioner and the Orchestrator. The former is used to interact with the cloud to acquire resources and the later is responsible for orchestrating the training processes.}
		\label{fig_arch}
	\end{figure}

	\begin{table}[htbp]
		\caption{User-Specified Parameters in SpotTune}
		\begin{center}
			\begin{tabular}{cp{0.3\textwidth}}
				\toprule
				$metric$&the metric used to evaluate an ML model's quality, e.g. validation loss, training MSE, etc.\\
				\midrule
				$max\_trial\_steps$& the maximum steps the user would like to train the ML model\\
				\midrule
				$\theta$&the early-shutdown rate to predict the final metric value using partial collected metric data\\
				\midrule
				$mcnt$&the number of models the user would like select from all the HPs\\
				\bottomrule
			\end{tabular}
			\label{tbl_param}
		\end{center}
	\end{table}

	In this section, we detail the design of SpotTune. SpotTune has two main concerns: \textbf{i)} finishing the HPT process with as low cost as possible, and \textbf{ii)} selecting (near) optimal models from all the HPs after a fixed number of iterating steps.
	
	\subsection{Overview}
	
	Figure \ref{fig_arch} depicts a high-level view of SpotTune and its operational environment. In a word, SpotTune takes advantage of cheap transient resources and the elasticity of the public cloud to do HPT in a cost-efficient manner. Noted that except for the ML model, all the HPs, and the training data, the user additionally needs to specify four other important parameters as described in Table \ref{tbl_param}, which could be assigned according to the real needs.

	
	\begin{algorithm}
		\caption{SpotTune Workflow}
		\label{algorithm_1}
		\begin{algorithmic}[1] 
			\Require $HPs$: hyper-parameter settings, $spotPool$: spot instance pool, $max\_trial\_steps$, $\theta$, $mcnt$
			\Function{getBestInst}{$t, hp, M$}
			\State $sCost \gets$ new Map()
			\For {$inst\ \in\ spotPool$}
			\State $price \gets \Call{random}{0.00001, 0.2}+price_{current}$
			\State $p \gets \Call{RevPredModel}{inst, t, price}$
			\State $sCost[inst] \gets M[inst][hp] \times (1-p) \times \overline{price}$
			\EndFor
			\State \Return{$\operatorname{argmin}{(sCost)}$}
			\EndFunction
			\State
			
			\State $assign \gets$ new Map($default\ value$=$null$)
			\State $M \gets$ new Map($default\ value$=$C_0 \times inst.CPU$)
			\State $waiting \gets \emptyset, finished \gets \emptyset$
			
			\State
			\While {True}
			\State $toQuery \gets (HPs-finished)$
			\If {$toQuery == \emptyset$}
			\State break
			\EndIf
			\For {$hp \in toQuery$}
			\State $VM \gets assign[hp]$
			\If {$VM$ == $null$}
			\State $waiting$.add($hp$)
			\ElsIf {receive the revocation notice of $VM$}
			\State checkpoint $result$ of $hp$ to S3
			\State $waiting$.add($hp$)
			\ElsIf {$hp$ has ran $\theta \times max\_trial\_steps$}
			\State checkpoint $result$ of $hp$ to S3
			\State shutdown $VM$
			\State $finished$.add($hp$)
			\ElsIf {$hp$ ran on $VM$ for $>$ one hour}
			\State checkpoint $result$ of $hp$ to S3
			\State shutdown $VM$
			\State $waiting$.add($hp$)
			\EndIf
			\State \Call{updateMetrics}{$M, inst(VM), hp$}
			\EndFor
			\For {$hp \in waiting$}
			\State $inst \gets \Call{getBestInst}{t_{now}}$
			\State $VM \gets \Call{requestAWS}{inst}$
			\State deploy $hp$ on $VM$
			\State $waiting$.remove($hp$)
			\State $assign[hp] \gets VM$
			\EndFor
			\State \Call{Sleep}{$10\ seconds$}
			\EndWhile
			\State
			\State $metrics \gets$ new Map()
			\For {$hp \in finished$}
			\State $metrics[hp] \gets \Call{EarlyCurve}{hp, max\_trial\_steps}$
			\EndFor
			\State \Call{sort}{$metrics$}
			\State continue training top-$mcnt$ models from checkpoints
		\end{algorithmic}
	\end{algorithm}
	
	\textbf{Provisioner} in SpotTune interacts with AWS to request and acquire resources. By computing the expected cost according to the output of RevPred and the online-updated performance data, Provisioner chooses the proper instances to deploy the HPT jobs.
	
	\textbf{Orchestrator} is responsible for monitoring the training trend and predicting the final metric of every HPT job using partially collected metrics (e.g., validation loss) during training. And when revocation happens, Orchestrator notifies the VM to checkpoint the intermediate training results to the remote object storage S3.
	
	Algorithm \ref{algorithm_1} describes the entire workflow of SpotTune to run the HPT jobs of a machine learning algorithm.
	
	The expected cost in the next hour for all the instances could be calculated using the following equation:
	\begin{equation}
	\begin{aligned}
		\mathbb{E}\left[eCost\right] &= [(1-p) \cdot \overline{price} + p \cdot 0] \cdot 1\ hour \\
		&= (1-p) \cdot \overline{price} \cdot 1\ hour
	\end{aligned}
	\end{equation}
	where $p$ is the revocation probability of current instance predicted by \textbf{RevPredModel} which we will introduce the detailed mechanism in the next subsection, and $\overline{price}$ is the average price of this instance in the last hour. Then function \textbf{getBestInst} returns the instance that has the lowest \textbf{\textit{step cost}} ($\$$ per step) in the next hour, which is the monetary cost of running a single iteration, defined as the following equation:
	\begin{equation}\label{equation_step_cost}
	\begin{aligned}
		\mathbb{E}\left[sCost\right] &= \mathbb{E}\left[\frac{M[inst][hp] \cdot eCost}{1\ hour}\right]  \\
		&=\frac{M[inst][hp] \cdot \mathbb{E}\left[eCost\right]}{1\ hour} \\
		&=\frac{M[inst][hp] \cdot (1-p) \cdot \overline{price} \cdot 1\ hour}{1\ hour} \\
		&=M[inst][hp] \cdot (1-p) \cdot \overline{price}
	\end{aligned}
	\end{equation}
	where $M$ is a two-dimensional matrix recording the performance of running different HPs on different instances. For example, $M[r3.xlarge][hp_1]$ (seconds per step) represents the speed of running HPT job $hp_1$ on instance $r3.xlarge$. $M$ is initiated according to the number of CPU cores of each instance. During the HPT process, $M$ would be updated in an online manner according to the latest runs as shown in line 36 of Algorithm \ref{algorithm_1}. This performance modeling method is practical because the computation pattern of a concrete HPT job is steady across different iterations, which we will demonstrate in our experiments (Subsection \ref{perf_profiling}).
	
	Then as the line 16-46 of Algorithm \ref{algorithm_1} shows, SpotTune loops to wait for three events: \textbf{1)} when receiving the pre-delivered revocation notice from AWS, SpotTune checkpoints its intermediate training results to S3 and arranges a re-deployment; \textbf{2)} when an HPT job finishes, SpotTune checkpoints its intermediate training results to S3 and marks it as $finished$; \textbf{3)} when an HPT job has run on an instance for more than one hour, SpotTune checkpoints its intermediate training results and shutdown the VM to arrange for a re-deployment.

	As we analyze in Section \ref{sec_intro}, HPT usually only leaves a small portion of models after trying them all, which means that a large amount of computation is meaningless. So SpotTune would not finish all the training steps, it would stop at $\theta \times max\_trial\_steps$ and predict the final quality (e.g., validation loss) of every model. After that, it would continue the training of the top-$mcnt$ models from the checkpoints. Line 48-53 reveals this procedure.
	
\begin{figure}[t]
		\centerline{\includegraphics[width=.5\textwidth]{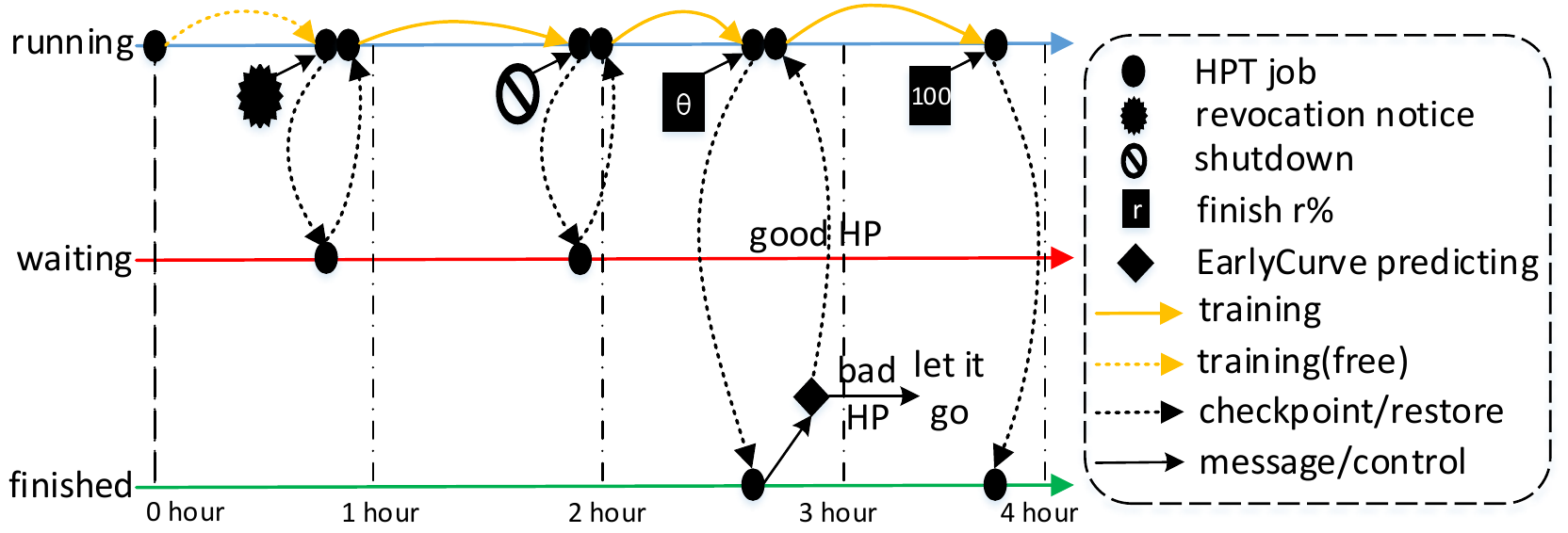}}
		\caption{An Example of the Lifetime of an HPT Job Using SpotTune.}
		\label{fig_workflow_example}
	\end{figure}

	Figure \ref{fig_workflow_example} depicts a typical example of the lifetime of an HPT job in SpotTune. As the demo figure shows, the VM that the model is deployed on is revoked in less than one hour so the initial tens of minutes are free. Then SpotTune forwardly shuts down the model training process when it has not been revoked for more than one hour. And when the model has been trained for $\theta \times max\_trial\_steps$, EarlyCurve determines whether to finish the entire training.
	
	According to Algorithm \ref{algorithm_1}, the functionality of SpotTune is guaranteed by two important prediction models, \textbf{RevPred} (line 5th) and \textbf{EarlyCurve} (line 50th). We show the details of these two models in the next two subsections.
	
	\subsection{Spot Market Price Modeling}\label{RevPred_info}
	
	As Equation \ref{equation_step_cost} shows, the revocation probability is one of the factors that SpotTune uses to calculate the \textbf{\textit{step cost}}. Give an instance type $I$, a \textit{maximum price} $b$ and a timestamp $t$, \textbf{RevPred} could output the revocation probability $P(I,b,t)$ of this instance in the next hour. The higher it is, the more possible SpotTune may select it. Similar to Tributary\cite{harlap2018tributary}, which is the state-of-the-art spot instance revocation probability predicting algorithm, RevPred uses an \textit{LSTM}-based deep neural network model structure to extract the history price knowledge. But in RevPred, there are two main differences: \textbf{i)} the \textit{LSTM} structure in RevPred only handles partial input data; \textbf{ii)} RevPred adopts a different data pre-processing approach to generate the \textit{maximum price}. For each individual spot market, an independent model is trained offline using the history spot price dataset.
	
	The input data of our model is separated into two parts. The first part is the history prices across the past 59 minutes, with each price record containing the following six engineered features: (1) current spot market price; (2) average spot market price; (3) the number of price changes in the past hour; (4) the time duration since the current spot market price is set; (5) whether the time is in the workdays or not; (6) current hour of the day. The second part is the present price record, with the above six features and the \textit{maximum price}.
	
	In Tributary, the \textit{maximum price} is randomly generated when training the LSTM model, which is slightly higher than the present spot market price by a random delta in a pre-defined interval ([0.00001, 0.2])\cite{harlap2018tributary}. While in RevPred, we argue that the delta should be determined differently in training and inference. When training the model, the \textit{maximum price} $b(I, t)$ at time $t$ is determined by the market price fluctuation as Algorithm \ref{algorithm_2}, aiming at calculating the average variation of instance $I$'s history prices (removing the smallest 20\% and the largest 20\%) in the previous 1 hours as the deltas in the training data. This idea comes from the essence of \textit{active learning}, which suggests that the data points locating on the classification border would help improve the model most significantly. And intuitively, the average variation of prices plus the current price would be close to the border that separates \textit{revoked} and \textit{not revoked}. If the spot market price grows over the maximum price in the following one hour, we label this record as \textit{True}. While using the trained model for inference, RevPred randomly generates the maximum price $b(I,t)$ as Tributary does.
	
	\begin{algorithm}
		\caption{Training Data Pre-processing}
		\label{algorithm_2}
		\begin{algorithmic}[1] 
			\Require $prices$: the history price data of spot instances used to train RevPred, $I$: spot instance type, $t$: timestamp
			\State $deltas \gets$ new List()
			\For {$(t-1\ hour) < \tau < t$}
			\State $deltas$.append($|prices[I][\tau] - prices[I][\tau-1]|$)
			\EndFor
			\State \Call{sort}{$deltas$}
			\State $L \gets \Call{length}{deltas}$, $b \gets 0$
			\For {$0.2 \times L < i < 0.8 \times L$}
			\State $b \gets b + deltas[i]$
			\EndFor
			\State $b \gets prices[I][t] + b / (0.6 \times L)$
		\end{algorithmic}
	\end{algorithm}
	
	The 59 price records in the past hour are fed into a three-tier \textit{LSTM} structure to retrieve the history spot market information, coming out as an embedding feature vector. The present price record is converted to an embedding presentation too through three sequential fully connected layers. Then these two feature vectors are concatenated together to come up with a probability-like result.
	
	The spot market price data could be very skewed. We use two techniques to mitigate the data-imbalance issue. First, we carefully design the loss function by assigning different weights for different classes. Let $\phi_+$ and $\phi_-$ denote the percentage of the positive samples (marked as \textit{True}) and negative samples in the training set, then we assign $\phi_-$ and $\phi_+$ as the positive and negative class weights. Second, we do not use the output of the prediction model as the probability directly. Denoting the output of the model for $I$ with price $b$ at time $t$ as $\hat{P}(I, b, t)$, we compute the preemption probability $P(I,b,t)$ as:
	\begin{equation}
		\frac{P(I,b,t)}{1-P(I,b,t)} = \frac{\hat{P}(I,b,t)\cdot\phi_-}{[1-\hat{P}(I,b,t)]\cdot\phi_+}
	\end{equation}
	Once a RevPred model is trained using the history price data of an instance $I$, it could predict the probability of preemption $P(I,b,t)$ in the following hour given the timestamp and the \textit{maximum price}.
	
	\subsection{ML Training Trend Modeling}\label{earlycurve_info}
	
	\begin{figure}[htbp]
		\centering
		\subfigure[Logistic Regression]{
			\label{fig_acc_curve}
			\includegraphics[width=0.45\columnwidth]{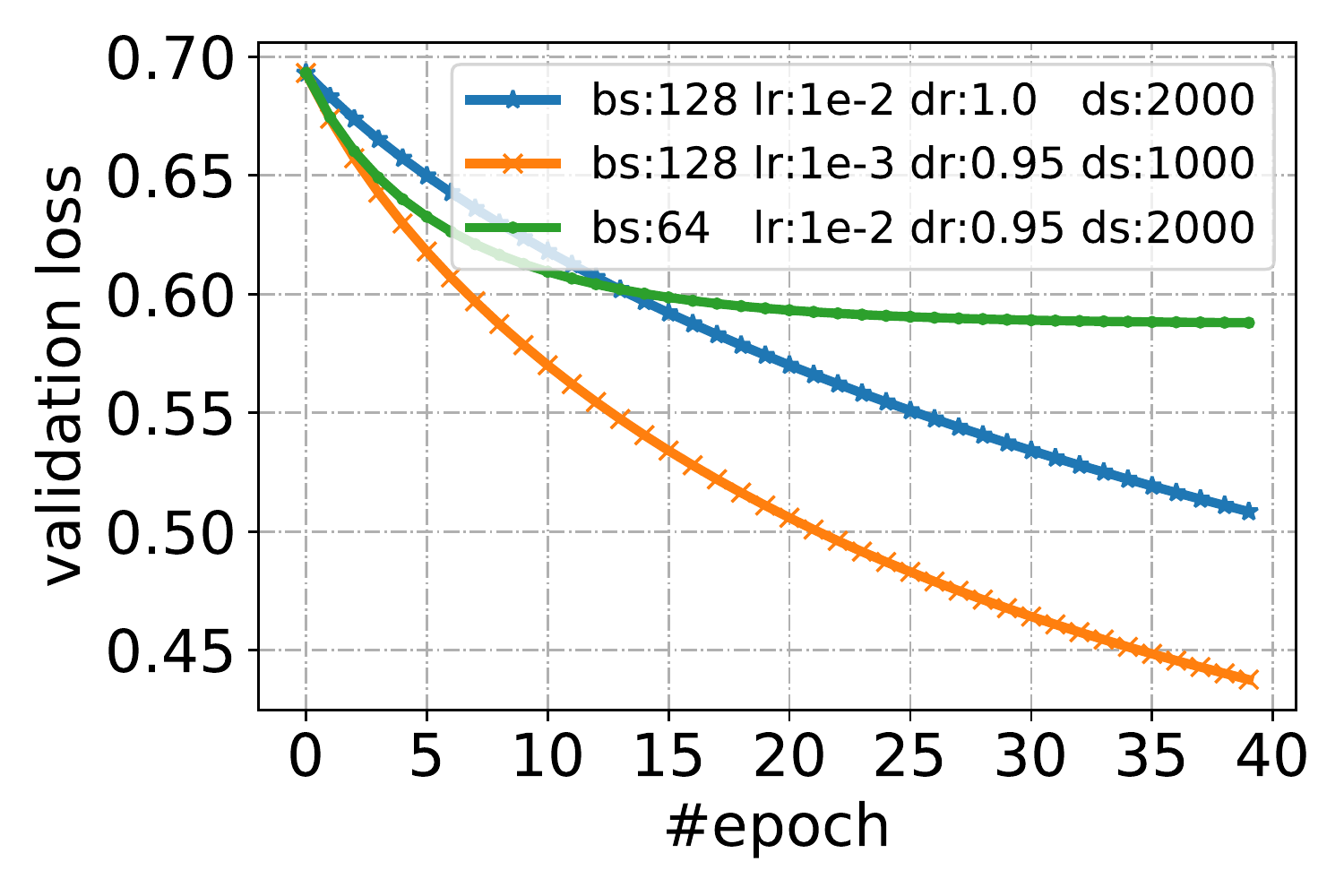}
		}
		\subfigure[Resnet-56]{
			\label{fig_res_loss}
			\includegraphics[width=0.45\columnwidth]{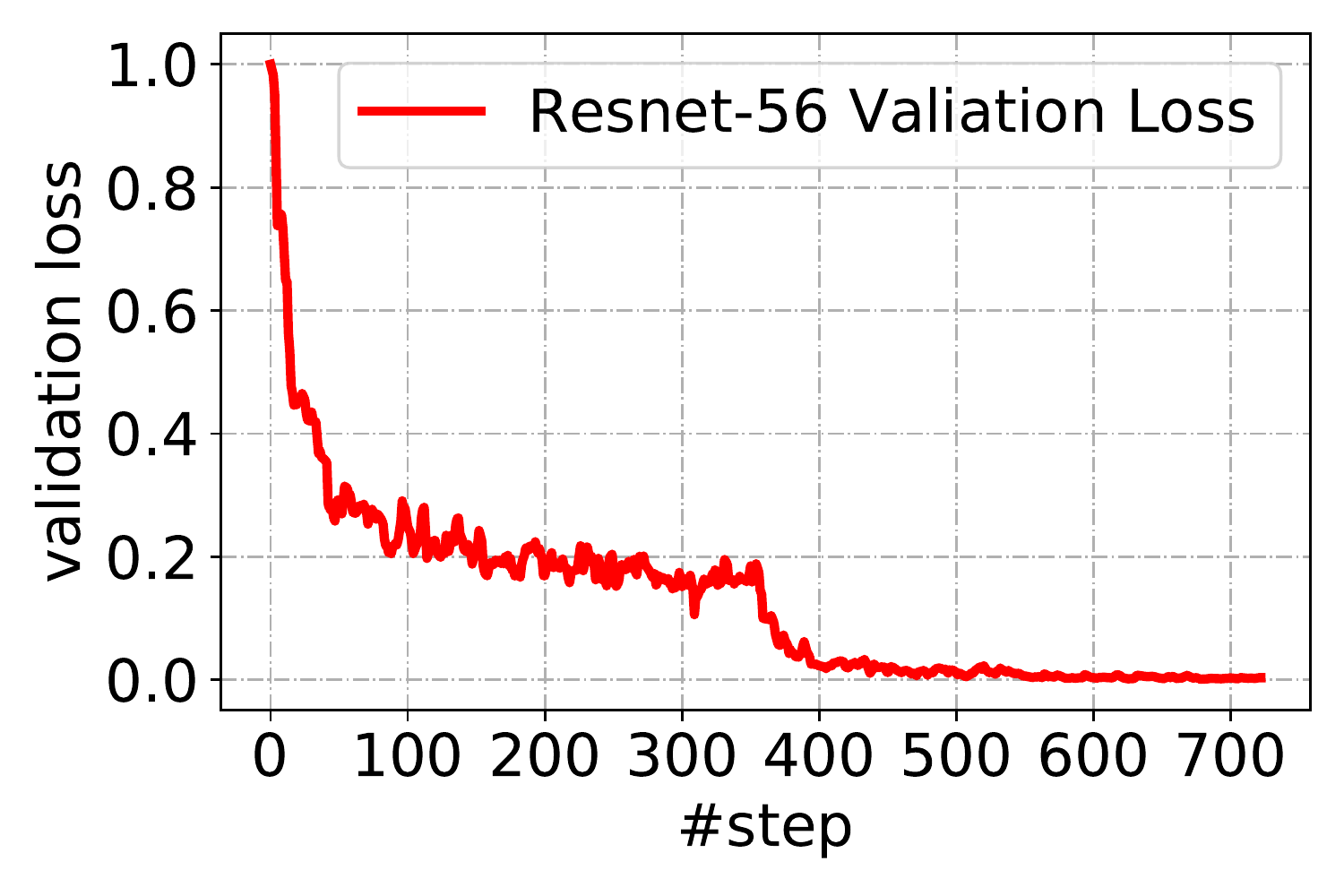}
		}
		\caption{(a): Logistic Regression validation loss(validating every epoch) curves on Epsilon dataset using three different hyper-parameter settings, where \textit{bs} denotes the batch size, \textit{lr} the initial learning rate, \textit{dr} the decay rate of the learning rate and \textit{ds} the number of steps between two learning-rate-decay. (b): Resnet-56 validation loss on CIFAR10 dataset. This model applies a learning rate that  decays periodically such that the validation curve is separated to two different stages.}
		\label{fig_curve}
	\end{figure}
	
	\newcommand{\tabincell}[2]{\begin{tabular}{@{}#1@{}}#2\end{tabular}}
	
	\begin{table*}
		\caption{Summary of ML algorithms, datasets, optimizers, metrics and hyper-parameter settings we used for testing. Notations: $bs$: batch size, $lr$: learning rate, $dr$: decay rate, $ds$: decay steps, $kernel$: SVM kernel function, $nt$: \#trees in GBTR, $version$: ResNet version, $depth$: max tree depth in GBTR or number of core covolutional layers in ResNet, $de$: decay epochs}
		\begin{center}
			\begin{tabular}{ccccc}
				\toprule
				\textbf{Algorithm}& \textbf{Dataset}& \textbf{Optimizer}& \textbf{Metric}& \textbf{HP settings}\\
				\midrule
				Logistic Regression(LoR)& Epsilon\cite{epsilon}& Gradient Descent& Cross Entropy& \tabincell{c}{$bs \in \{128,64\}$
					$lr \in \{1e-2,1e-3\}$ \\
					$dr \in \{1.0,0.95\}$
					$ds \in \{1000,2000\}$} \\
				\midrule
				Support Vector Machine(SVM)& Synthetic& Gradient Descent& Hinge Loss& \tabincell{c}{$bs \in \{128,64\}$
					$lr \in \{1e-2,1e-3\}$ \\
					$dr \in \{1.0,0.95\}$ \\
					$kernel \in \{RBF,Linear\}$} \\
				\midrule
				GBT Regression(GBTR)& Synthetic& Gradient Boosting& MSE& \tabincell{c}{$bs \in \{128,64\}$
					$lr \in \{1e-1,1e-2\}$ \\
					$nt \in \{10,15\}$
					$depth \in \{5,8\}$} \\
				\midrule
				Linear Regression(LiR)& YearPredictionMSD\cite{yearpred}& Gradient Descent& MSE& \tabincell{c}{$bs \in \{128,64\}$
					$lr \in \{1e-2,1e-3\}$ \\
					$dr \in \{1.0,0.95\}$
					$ds \in \{1000,2000\}$} \\
				\midrule
				AlexNet(AlexNet)& Cifar10\cite{cifar10}& Adam\cite{kingma2014adam}&Cross Entropy& \tabincell{c}{$bs \in \{128,64\}$
					$lr \in \{1e-1,1e-2\}$ \\
					$dr \in \{1.0,0.95\}$
					$de \in \{40,60\}$} \\
				\midrule
				Residual Neural Network(ResNet)& Cifar10& Adam&Cross Entropy& \tabincell{c}{$bs \in \{32,64\}$
					$version \in \{1,2\}$ \\
					$depth \in \{20,29\}$
					$de \in \{40,60\}$} \\
				\bottomrule
			\end{tabular}
			\label{tbl_bench}
		\end{center}
	\end{table*}
	
	\textbf{EarlyCurve} uses partial metrics to predict the final metric of an ML model, thus to save computation capacity by shutting down bad models early. The user could indicate different metrics for different problems. For example, training/validating cross-entropy for classification problems and training/validating MSE (mean squared error) for regression problems. The algorithms and metrics used in our experiments are shown in Table \ref{tbl_bench}. ML algorithms listed in Table \ref{tbl_bench} use gradient-based optimization approaches to update trainable parameters, which converges (assume the loss function $f$ is convex, differentiable, and $\nabla f$ is Lipschitz continuous) at a rate of O($\frac{1}{k}$) in terms of the number of steps \textit{k} (with optimized versions of gradient descent/boosting, the convergence rate could be improved to O($\frac{1}{k^2}$))\cite{peng2018optimus}. Figure \ref{fig_acc_curve} shows an example of different shapes of validation loss curves using different HP settings. EarlyCurve uses the following model (algorithms optimized with optimizers that converge at a different rate are discussed in Section \ref{section_6}) to fit the validation metrics during the training process:
	\begin{equation}\label{equation_acc}
	\hat{\mathcal{L}_k} = \sum_{i=1}^{ST}(\frac{1}{\alpha_{i0} \cdot k^2 + \alpha_{i1} \cdot k + \alpha_{i2}} + \alpha_{i3}) \cdot sign(k, l_i, r_i)
	\end{equation}
	where $\hat{\mathcal{L}_k}$ denotes the model's predicted metric at step (epoch) $k$, $\alpha_{i0}$, $\alpha_{i1}$, $\alpha_{i2}$ and $\alpha_{i3}$ are non-negative coefficients, and function $sign$ is defined as: 
	\begin{equation}
	sign(k, l_i, r_i) = 
	\left\{
	\begin{array}{lc}
	1, & l_i \leq k \leq r_i\\
	0, & k < l_i\ or\ k > r_i
	\end{array}
	\right.
	\end{equation}
	Where $l_i$ and $r_i$ denote the left and right bound of an interval, which satisfies the following condition at time $T$:
	\begin{equation}
	\begin{aligned}
		\bigcup_1^T[l_i, r_i) = [0, T),\ and,\ \bigcap_i^T[l_i, r_i) = \emptyset
	\end{aligned}
	\end{equation}
	Unlike prior works\cite{peng2018optimus,zhang2017slaq}, Equation \ref{equation_acc} takes the multi-stage issue like Figure \ref{fig_res_loss} into consideration. That is, in the training process of an ML model, the learning rate could be different at different stages ($ST$ denotes the \#stage) such that the metrics could vary sharply at some certain points, so we model the training process as a piecewise function. EarlyCurve fits the training curve in an online manner, determining when to separate the curve to a new stage according to the following criteria:
	\begin{equation}\label{equation_new_stage}
	\begin{aligned}
	&\zeta_i = \frac{|\mathcal{L}_i - \mathcal{L}_{i-1}|}{\mathcal{L}_{i-1}} > \xi, \\
&	\zeta_j < \varepsilon, j \in \{i-1,...,i-5\} 
	\end{aligned}
	\end{equation}
	where $\mathcal{L}_i$ denotes the to-be-evaluated metric at step $i$, $\zeta_i$ the changing rate of $\mathcal{L}$ at step $i$, $\xi$ the pre-defined threshold of metric's changing rate (0.5 in our experiment), and $\varepsilon$ the threshold of last five steps' change rate (0.01 in our experiment). Equation \ref{equation_new_stage} is heuristic: if the changing rate of a model's metric is suddenly high after a steady period, it could be considered to be moving to a new stage.
	
	The least number of steps used to predict the final metric is controlled by the parameter $\theta$. For example, if $\theta$=0.8 and $max\_trial\_steps$=10000, EarlyCurve would predict the final metric after running 8000 steps. Through our experiments, we observe that the starting 70\% (see Section \ref{section_5} for more detail) points of all the training steps could support a well-behaved fitting model, mostly predicting a nearly 70\% accuracy and 100\% top-3 accuracy. We measure SpotTune's sensitivity against $\theta$ in Section \ref{section_5}.
	
	One extra special case should be mentioned, which is that the model comes to convergence before $\theta \cdot max\_trial\_steps$ steps. After that, the metric curve becomes a plateau, where training is no longer meaningful anymore. In this case, we stop the iteration and treat this model as finished.
	
	EarlyCurve works as follows: As the training progress goes on, the SpotTune Orchestrator keeps monitoring the metric data point ($k$, $\mathcal{L}$). If the number of already run steps is above $\theta \times max\_trial\_steps$, EarlyCurve then uses a linear regression solver\cite{nnls} to find the best coefficients that fit the collected metric data points and predict the final metric value.
	
	\section{Evaluation}\label{section_5}
	
	We evaluate SpotTune from four aspects. First, we evaluate the cost-saving effects and performance by simulating the HPT process using real-world ML benchmarks and EC2 spot market history price data. Second, we try to explain why SpotTune is cost-effective. Third, we analyze the performance of RevPred and EarlyCurve compared with previous works. Finally, we discuss the overhead of our system.
	
	\subsection{Experimental Setup}
	
	\begin{table}
		\caption{Experimental Instance Configurations}
		\begin{center}
			\begin{tabular}{cccc}
				\toprule 
				Instance Type& CPUs& Memory(GB)& On-demand Price(USD/hr)\\
				\midrule
				\textit{r4.large}& 2& 15.25& 0.133\\
				\textit{r3.xlarge}& 4& 30& 0.33\\
				\textit{r4.xlarge}& 4& 30.5& 0.266\\
				\textit{m4.2xlarge}& 8& 32& 0.4\\
				\textit{r4.2xlarge}& 8& 61& 0.532\\
				\textit{m4.4xlarge}& 16& 64& 0.8\\
				\bottomrule
			\end{tabular}
			\label{tbl_conf}
		\end{center}
	\end{table}

	\subsubsection{\textbf{Spot Market Price Data}} We use the open-source dataset \textit{AWS Spot Pricing Market}\cite{kaggledata} publicized on Kaggle. It collects the spot prices across eleven regions from 2017-04-26 to 2017-05-08. Theoretically, any price datasets could be adopted in the training of RevPred. We use the \textit{us-east-1.csv} sub-dataset, covering 68 types of instances. For simplicity, we choose six representative types as the experimental spot instance pool. The configurations and on-demand prices of these five instances are shown in Table \ref{tbl_conf}. Noted that this dataset is rather sparse, which means that the time intervals between adjacent records are non-uniform. So we preprocess the dataset by interpolating values between records, making the timestamp interval between adjacent records fixed at 1 minute.
	
	\subsubsection{\textbf{ML benchmarks}} We test our system with some popular ML algorithms, including \textbf{(i)} classification algorithms: Logistic Regression, SVM, and CNN-based neural network schema AlexNet and ResNet; \textbf{(ii)} regression algorithms: Gradient Boosting Tree Regression and Linear Regression. For each algorithm, we try a set of different hyper-parameter settings as shown in Table \ref{tbl_bench}. For example, SVM with different kernels, ResNet with different numbers of core convolutional layers.
	
	\subsubsection{\textbf{Datasets}} To evaluate SpotTune's general ability in the real world, we use multiple datasets we collected from various online sources with some modifications and our own dataset. The datasets span a variety of types (including images\cite{mnist, cifar10}, binary-classification bench\cite{epsilon}, audio meta-features\cite{yearpred}, and our random-generated synthetic dataset), as shown in Table \ref{tbl_bench}. In our experiments, all the training datasets are serialized to python pickle objects, summing up to over 100GB.
	
	\subsubsection{\textbf{Baseline}} 
\begin{figure*}[htbp]
		\centering
		\subfigure[Overall Cost.]{
			\label{fig_overall_cost}
			\includegraphics[width=0.64\columnwidth]{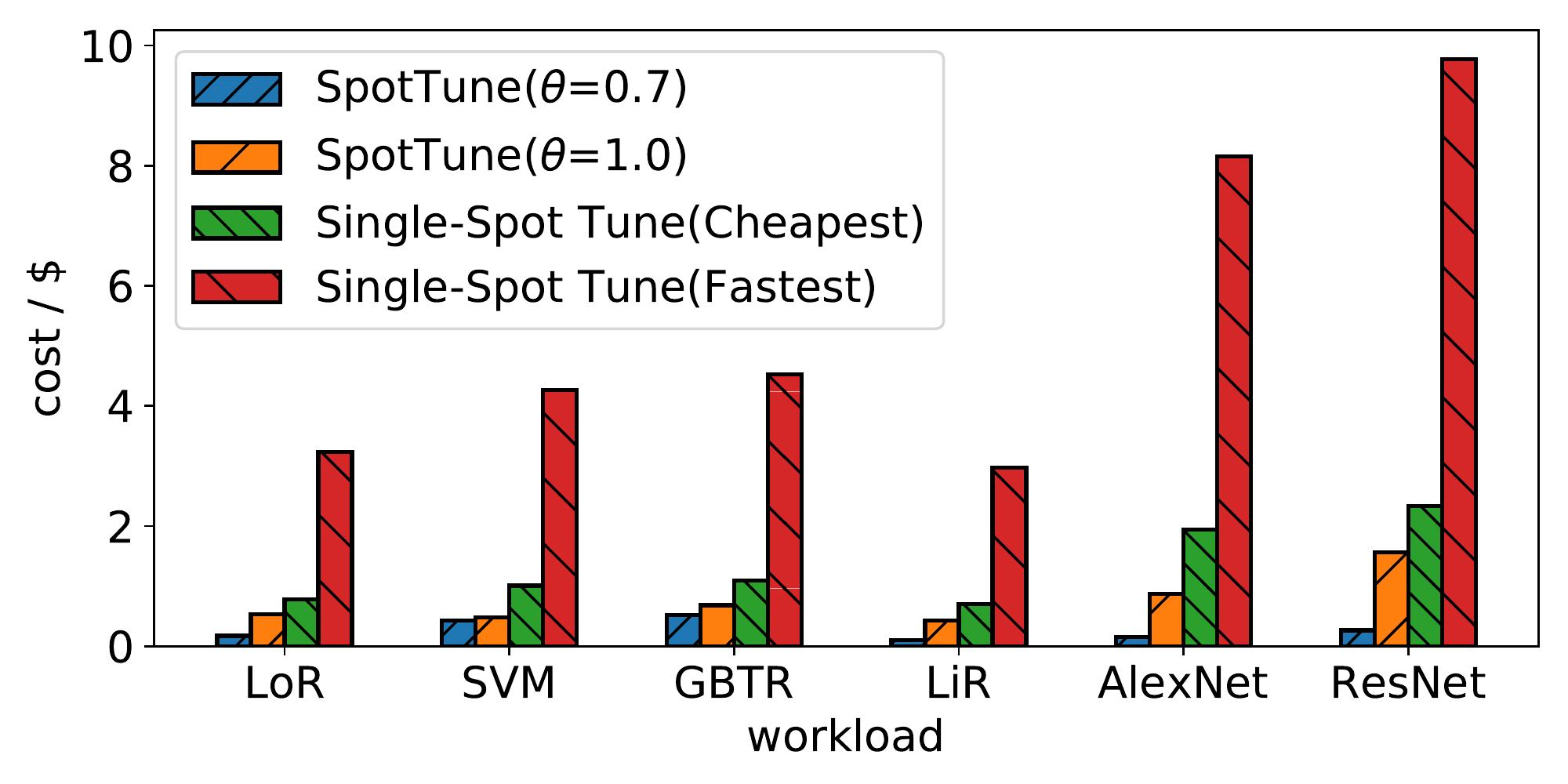}
		}
		\subfigure[Job Completion Time.]{
			\label{fig_fin_time}
			\includegraphics[width=0.64\columnwidth]{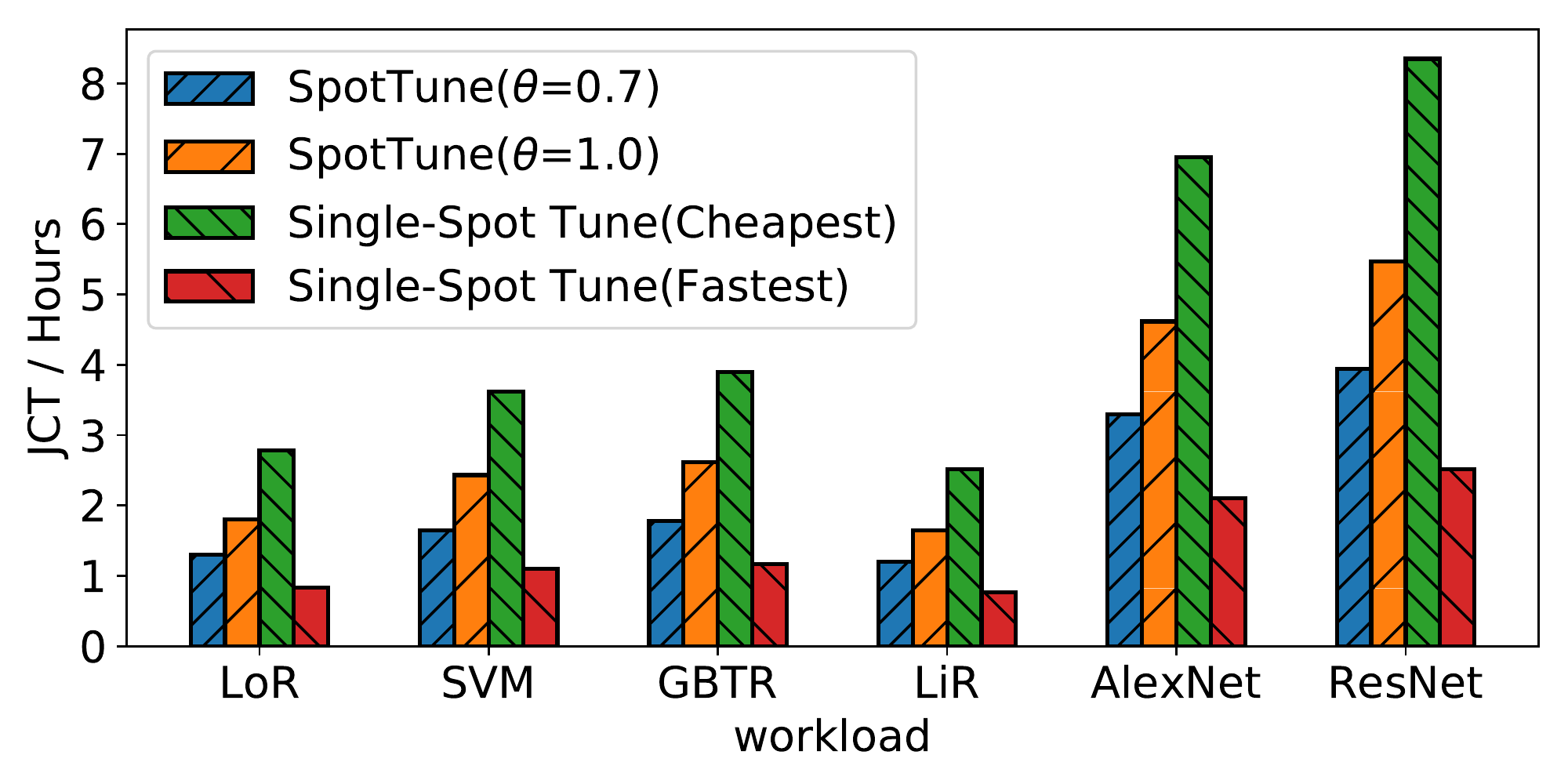}
		}
		\subfigure[Performance-Cost Rate.]{
			\label{fig_pcr}
			\includegraphics[width=0.64\columnwidth]{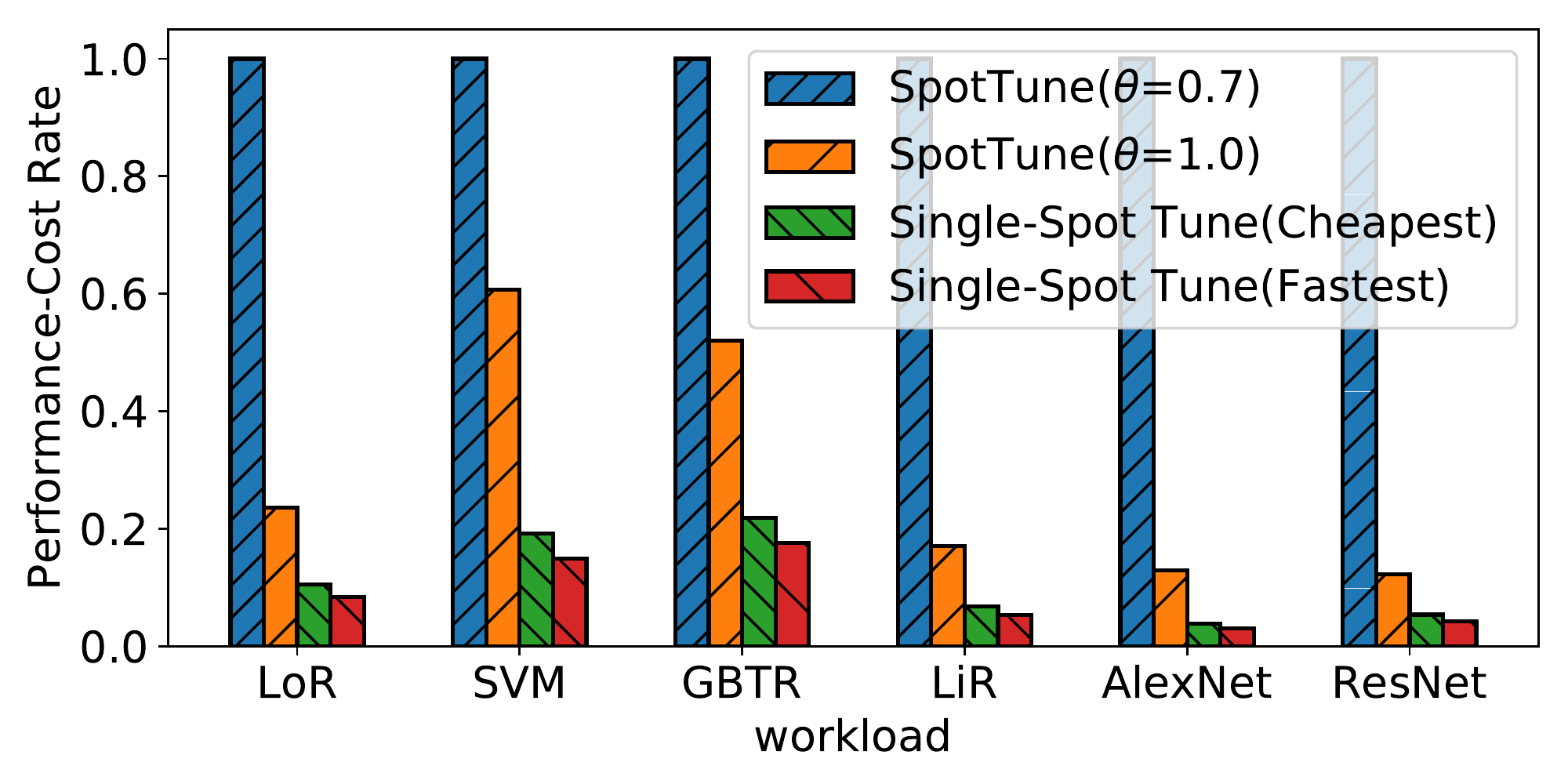}
		}
		\caption{Cost/Performance Comparison between SpotTune and Two Baselines.}
		\label{fig_overall}
	\end{figure*}

The baseline we compare SpotTune with is running HPT on a single spot instance. We assume the \textit{maximum price} of each used single-spot instance is much higher than its market price such that it would not be revoked. Many previous works adopt similar strategies to mitigate the risk of revocation\cite{subramanya2015spoton}. There are two kinds of representative spot instances in our experiment, the \textit{cheapest} (\textit{r4.large}) and the \textit{fastest} (\textit{m4.4xlarge}), denoted as \textit{Single-Spot Tune (Cheapest)} and \textit{Single-Spot Tune (Fastest)} in Figure \ref{fig_overall}. There are two version SpotTune in our experiment, with $\theta$ set as 0.7 and 1.0 respectively. As the Top-3 accuracy reaches 100\% when $\theta \geq 0.7$, we treat 0.7 as the minimum reliable $\theta$ by default. If the user indicates a $\theta$ of less than 0.7, he might take the misprediction risk of EarlyCurve.
	
	\subsubsection{\textbf{Performance Profiling}}\label{perf_profiling}
	\begin{figure}[htbp]
		\centerline{\includegraphics[width=.45\textwidth]{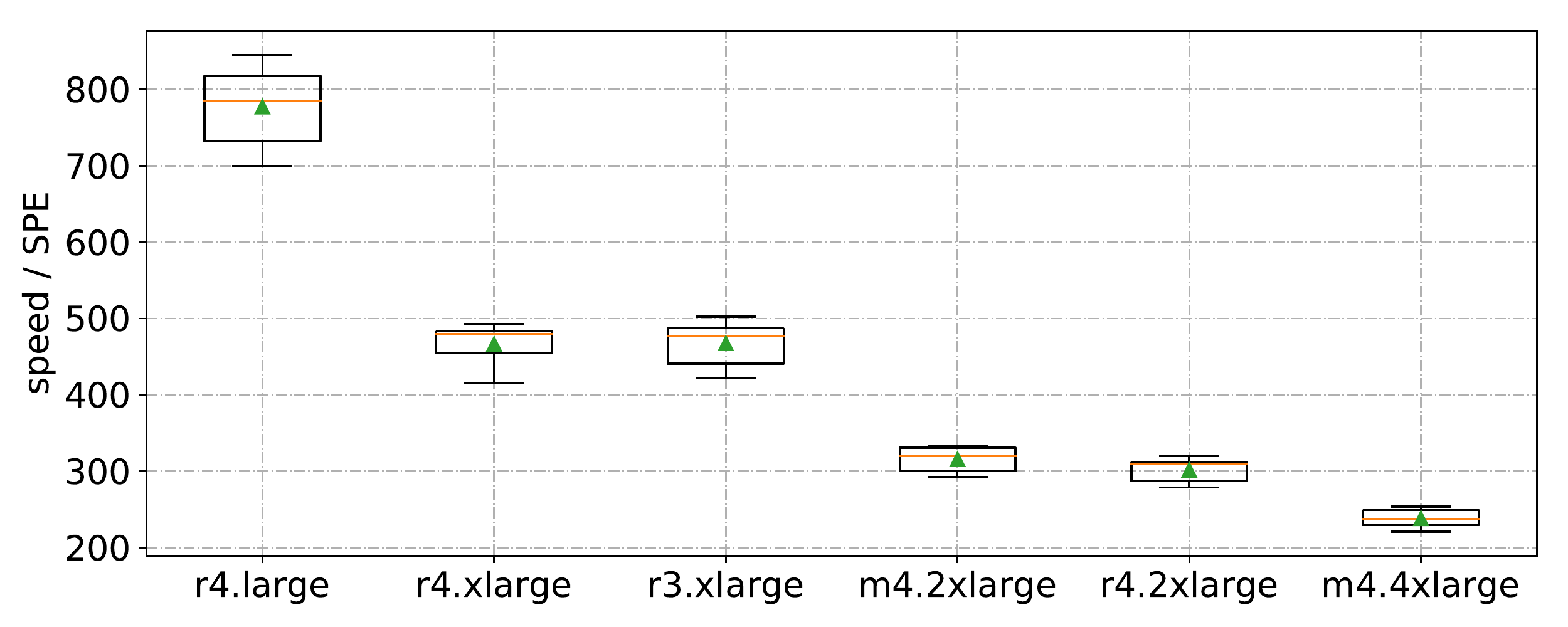}}
		\caption{Performance Profiling Example.}
		\label{fig_perf_profiling}
	\end{figure}

	\begin{figure*}[htbp]
		\centering
		\subfigure[Cost.]{
			\label{fig_cost_theta}
			\includegraphics[width=0.64\columnwidth]{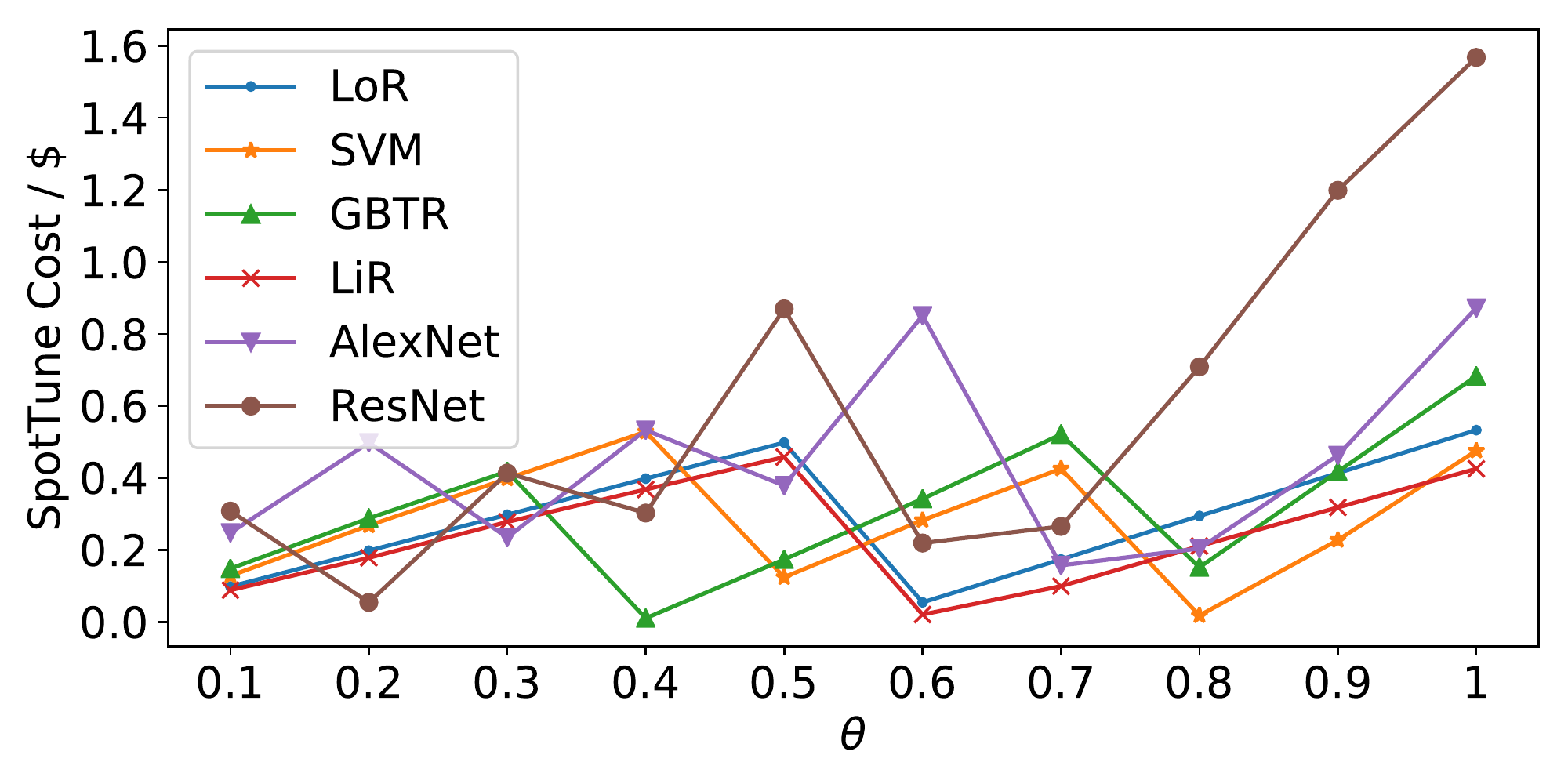}
		}
		\subfigure[Job Completion Time.]{
			\label{fig_fin_time_theta}
			\includegraphics[width=0.64\columnwidth]{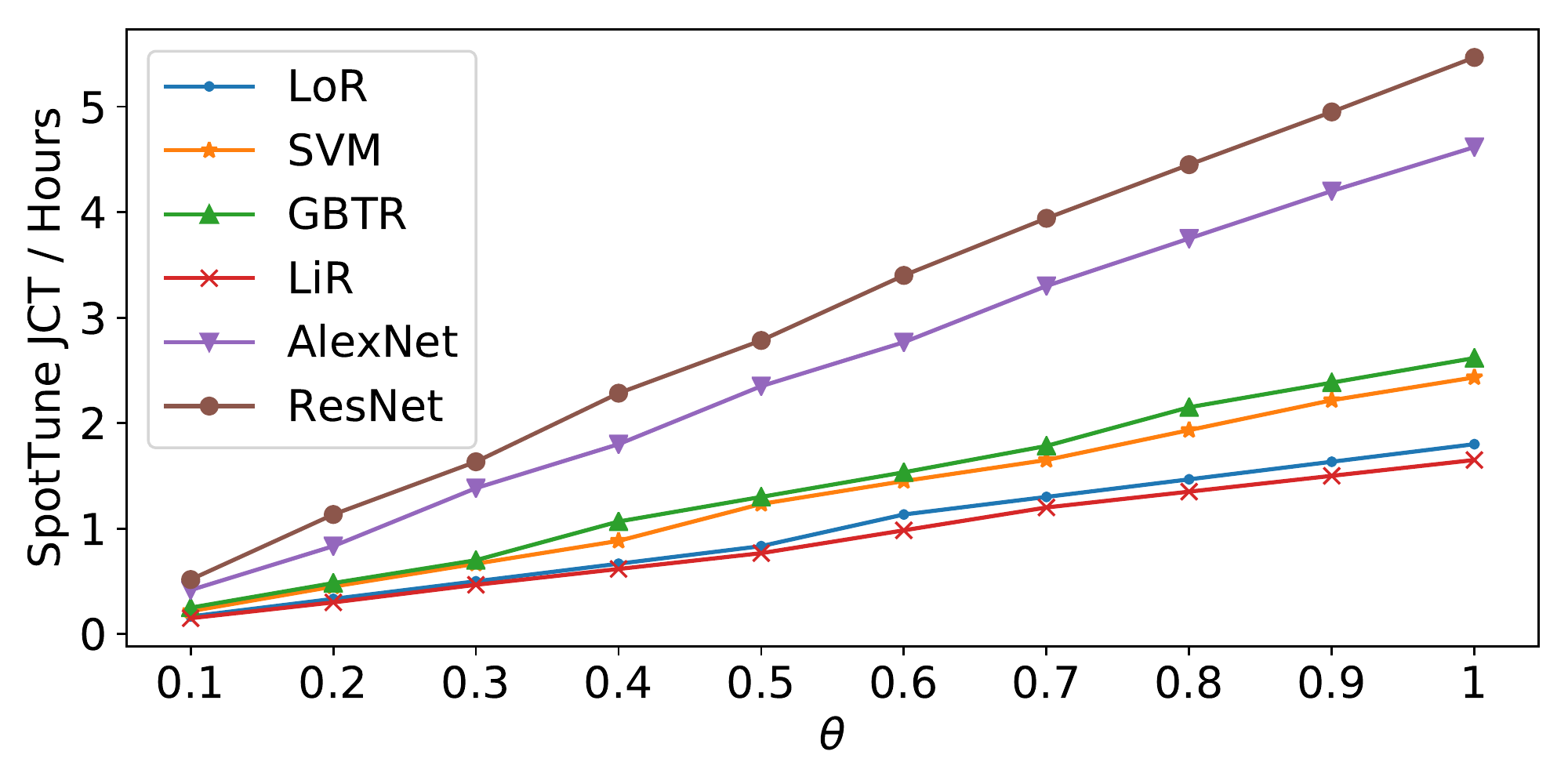}
		}
		\subfigure[Accuracy.]{
			\label{fig_top3_acc_theta}
			\includegraphics[width=0.64\columnwidth]{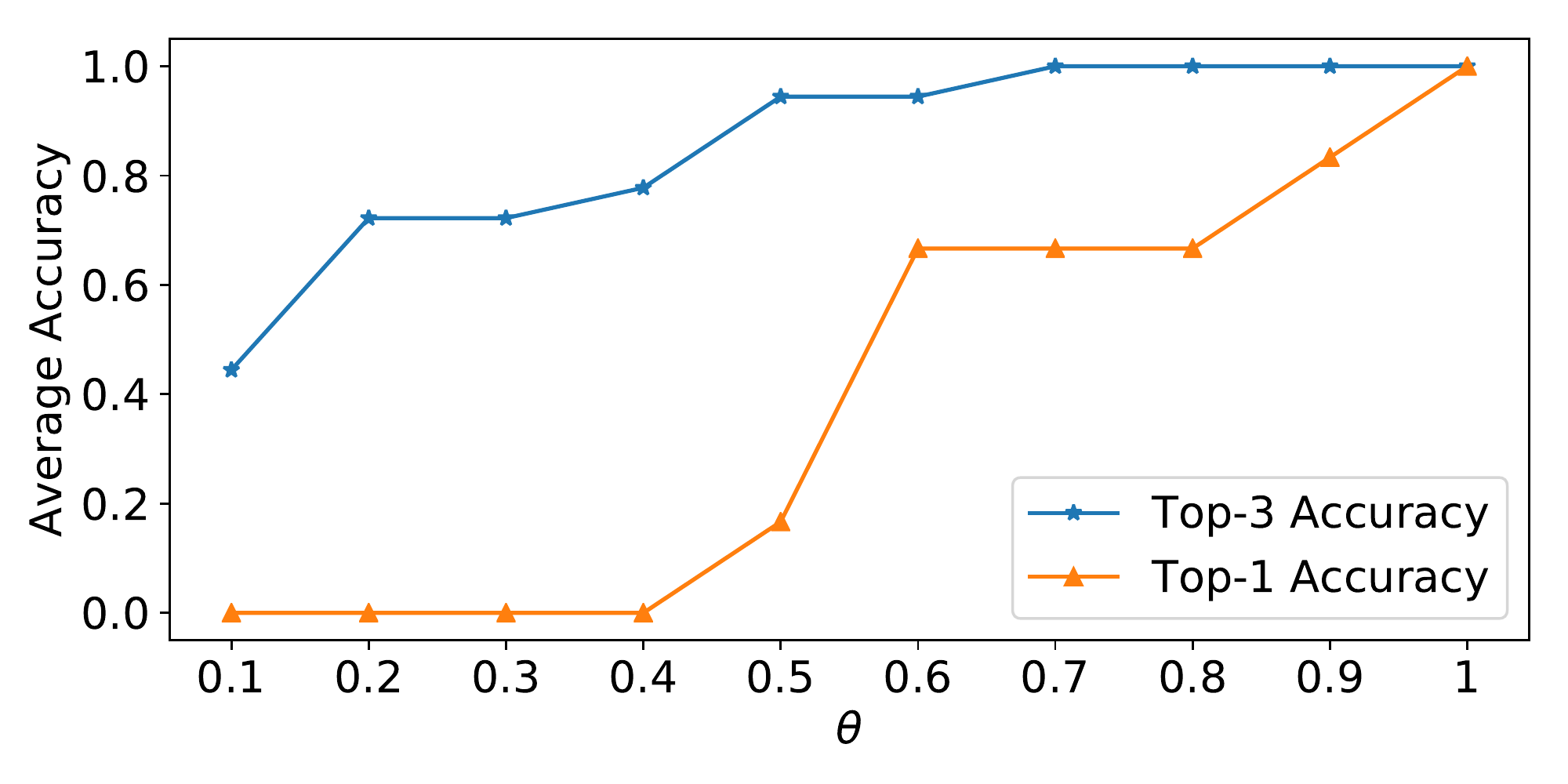}
		}
		\caption{SpotTune's Sensitivity against $\theta$}
		\label{fig_theta}
	\end{figure*}
	As described in Algorithm \ref{algorithm_1}, we profile the performance of running different HPT jobs on different instances in an online manner. To prove the feasibility of this profiling method, we pre-run all the models in Table \ref{tbl_bench} on all the instances in Table \ref{tbl_conf} for $N$ steps (epochs) and record the time consumption of each step (epoch) $seconds/batch (SPB)$ (or $seconds/epoch (SPE)$) as $M[inst][HP]$. We observe that the coefficient of variation (COV) of $M[inst][HP]$ across different steps is small enough (less than 0.1), which demonstrates the online profiling method is practical. Figure \ref{fig_perf_profiling} shows an example of the performance metrics of different types of spot instances in our experiment, whose vertical axis represents the speed (SPE) of training the ResNet model. And as the horizontal axis goes right, the price of the instance is increasing. An interesting observation goes to that as the prices increases, the performance does not get better accordingly all the time, at least not in a linear manner.
	
	\subsection{Cost and Performance Evaluation}
	
	\subsubsection{\textbf{Overall Cost and Performance}}

	To evaluate the cost-saving effect and performance of SpotTune, we simulate the HPT process of all the benchmark workloads shown in Table \ref{tbl_bench}. Figure \ref{fig_overall} shows the simulation results. When $\theta$ is set 1.0, the EarlyCurve component does not work, so the two baselines could be considered as $\theta$=1.0 as well. It should be mentioned again that we assume the \textit{maximum price} of the single-spot instance approach is high enough to avoid the preemption.
	
	In Figure \ref{fig_overall_cost} we show the overall cost yielded by four approaches. Averagely speaking, the overall cost of \textit{SpotTune ($\theta$=0.7)} is the lowest across all the six simulating workloads. \textit{SpotTune ($\theta$=1.0)} saves 41.5\% and 86.04\% compared with the cheapest and the fastest standalone spot instance respectively. Besides, \textit{SpotTune ($\theta$=0.7)} saves 57.16\% compared with \textit{SpotTune ($\theta$=1.0)} because of the 30\% less running steps. Finally, \textit{SpotTune ($\theta$=0.7)} could save up to 75.64\% and 94.18\% compared with \textit{Single-Spot Tune (Cheapest)} and \textit{Single-Spot Tune (Fastest)} because of the refund bonus and the less trying steps.
	
	Figure \ref{fig_fin_time} illustrates the HPT job completion time (JCT) of four approaches, where JCT is defined as the time span from the HPT job submission to selecting the best model(s). JCT composes of two main parts: training and checkpoint-restore. We will evaluate the checkpoint-restore time percentage in Subsection \ref{sys_overhead}. Intuitively, the JCT of SpotTune is between two baselines because it uses a mixture of all the instances. \textit{SpotTune ($\theta$=0.7)} can achieve 64.59\% of the fastest instance's performance and have a 2.14x boosting compared with the cheapest instance.
		
	Figure \ref{fig_pcr} depicts the performance-cost rate (PCR) of four approaches, which is calculated by $\alpha / (JCT \cdot cost)$ where $\alpha$ is a constant, revealing each approach's performance of a unit cost. The less the JCT and the cost are, the higher the performance-cost rate is. We normalize the PCR data for each benchmark, fixing the \textit{SpotTune ($\theta$=0.7)} PCR as 1. With all running the same number of steps, \textit{SpotTune ($\theta$=1.0)} has a 2.65x and 3.36x PCR compared with \textit{Single-Spot Tune (Cheapest)} and \textit{Single-Spot Tune (Fastest)}. With $\theta$ set as 0.7, \textit{SpotTune ($\theta$=0.7)} has a 13.11x and 16.61x PCR compared with the two baselines.
			
	\subsubsection{\textbf{Sensitivity against $\theta$}}
	
	SpotTune has one parameter $\theta$ that needs to be determined by the user, so we evaluate the sensitivity of SpotTune against $\theta$. Figure \ref{fig_theta} shows the results. $\theta$ reflects the user's attitude toward early-shutdown. If $\theta$ is small (e.g., 0.1), it reflects that the user is radical so as to select the \textit{best} model more quickly yet might take the risk of misprediction. Otherwise, the user may want to do the HPT in a safe and sound manner. We let the $\theta$ vary from 0.1 to 1.0 with an interval of 0.1.
	
	Figure \ref{fig_cost_theta} depicts the cost yielded by SpotTune under different $\theta$ and benchmarks. Generally speaking, the overall cost is proportional to $\theta$. But sometimes the higher $\theta$ yields lower cost instead. Taking the \textit{SVM} benchmark as an example, the overall cost when $\theta$=0.8 is less than when $\theta$=0.7. This is because some instances are terminated by SpotTune when the already-run steps achieving $\theta \times max\_trial\_steps$ when $\theta$ is lower but are revoked by AWS when $\theta$ is higher thus incurring more refund. The JCT of SpotTune varies near-linear to $\theta$ as illustrated in Figure \ref{fig_fin_time_theta}. Figure \ref{fig_top3_acc_theta} shows the accuracy/top-3 accuracy of the final selected model, i.e., the accuracy of EarlyCurve. When $\theta$ comes to 0.7 or higher, EarlyCurve can predict a 100\% top-3 accuracy.
	
	According to SpotTune's sensitivity against $\theta$, the user could indicate $\theta$ to meet his demand. For example, if the user wants to select the \textit{near-best} model as quickly as possible, he could indicate the $\theta$ as 0.2-0.4, getting a 70\%-80\% top-3 accuracy. If the user persists to get the highest top-1 accuracy, $\theta$ should be assigned as 1.0, i.e., without early-shutdown.
	
	\subsection{Why SpotTune is the Cheapest}\label{why_cheapest}

\begin{figure}[htbp]
		\centering
		\subfigure[Free Resources Contribution.]{
			\label{fig_free_steps}
			\includegraphics[width=0.45\columnwidth]{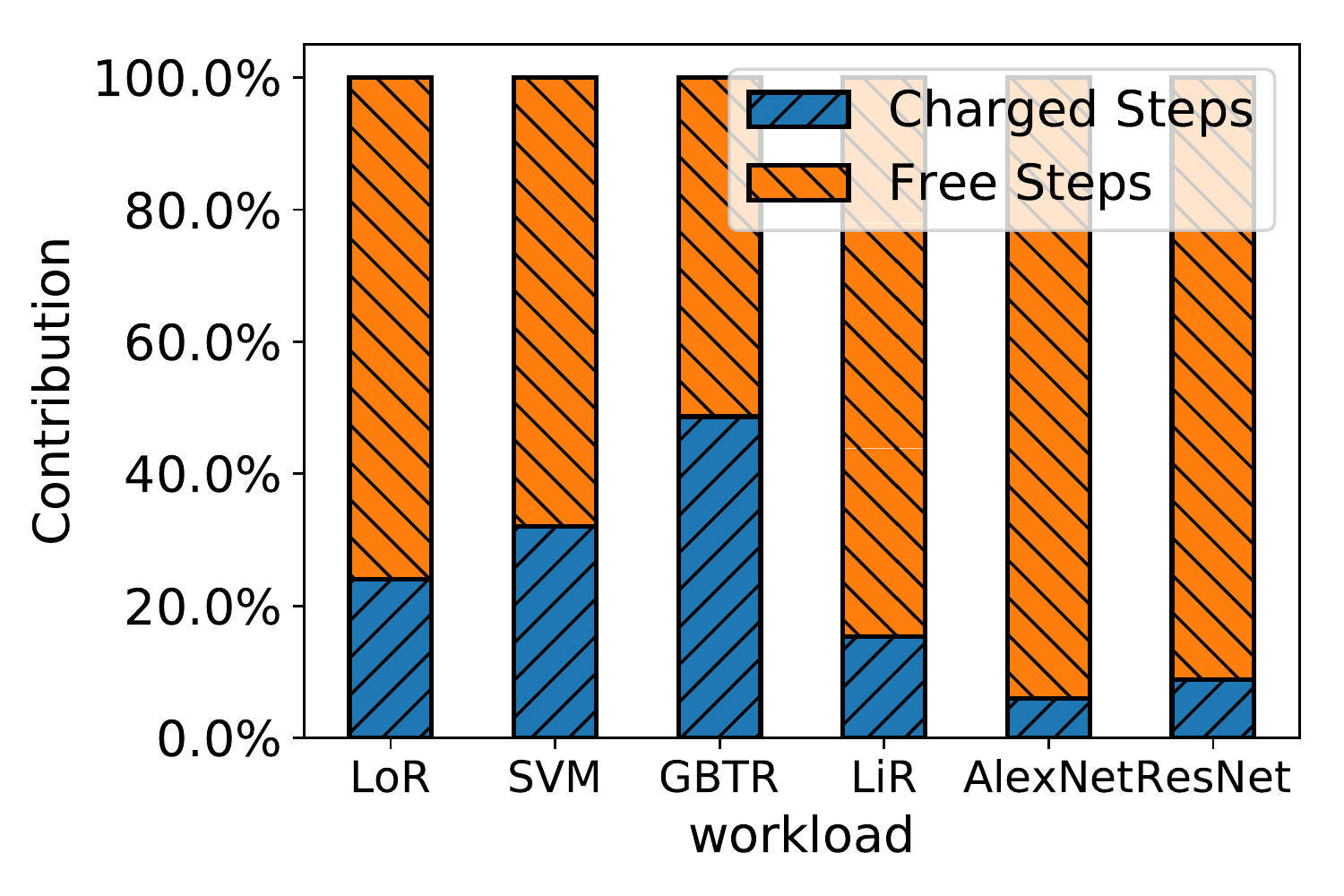}
		}
		\subfigure[Refund-Cost Comparison.]{
			\label{fig_free_cost}
			\includegraphics[width=0.45\columnwidth]{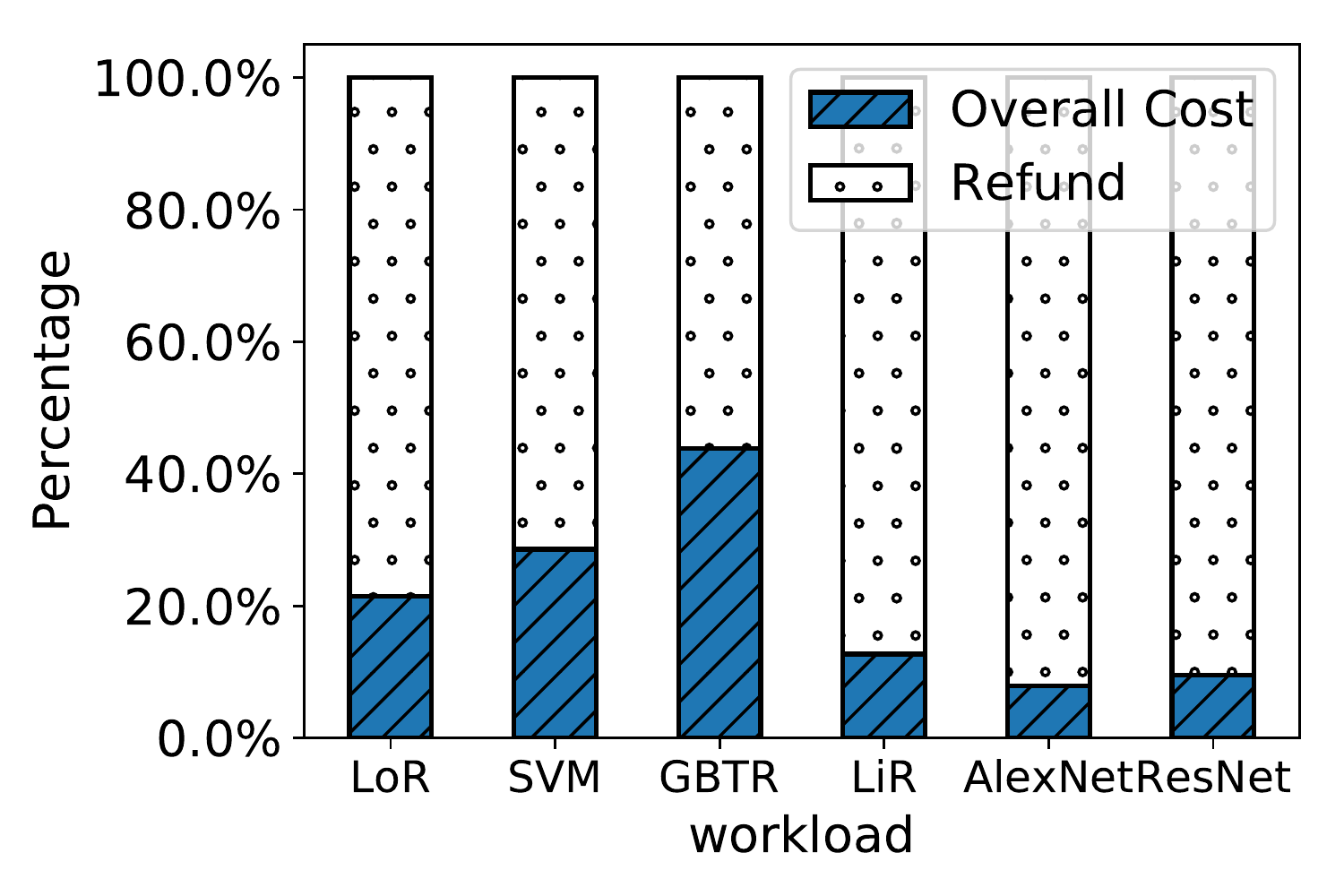}
		}
		\caption{Contribution/Cost-saving Percentage of the Refunded Resources.}
		\label{fig_free}
	\end{figure}

	Figure \ref{fig_overall} shows SpotTune is faster yet cheaper than the single cheapest spot instance. We find that there are two main reasons: \textbf{i)} the instance with the lowest price may not be the cheapest instance to run the entire workload; \textbf{ii)} the minimization of Equation \ref{equation_step_cost} relies on three factors: performance average price, and revocation probability of an instance. As we always re-schedule the running instance every one hour, the instance with higher revocation probability would be favored by SpotTune. We evaluate the numeric benefits brought by the refunded resources. Figure \ref{fig_free} shows the contribution/cost-saving percentage of the refunded (free) resources. As shown in Figure \ref{fig_free} ($\theta=0.7$), the refunded resources contribute an average of 77.5\% steps. In other words, taking the Logistic Regression benchmark in which we set the $max\_trial\_steps$ as 1000 as an example, there are $1000 \times 0.7 \times 75\%=525$ steps are run freely. The refund is an important reason that SpotTune is faster yet cheaper than \textit{Single-Spot Tune (cheapest)}.
	
	\subsection{RevPred Performance}\label{RevPred_perf}
	
\begin{figure}[!h]
		\centering
		\subfigure[Accuracy.]{
			\label{fig_pred_acc}
			\includegraphics[width=0.45\columnwidth]{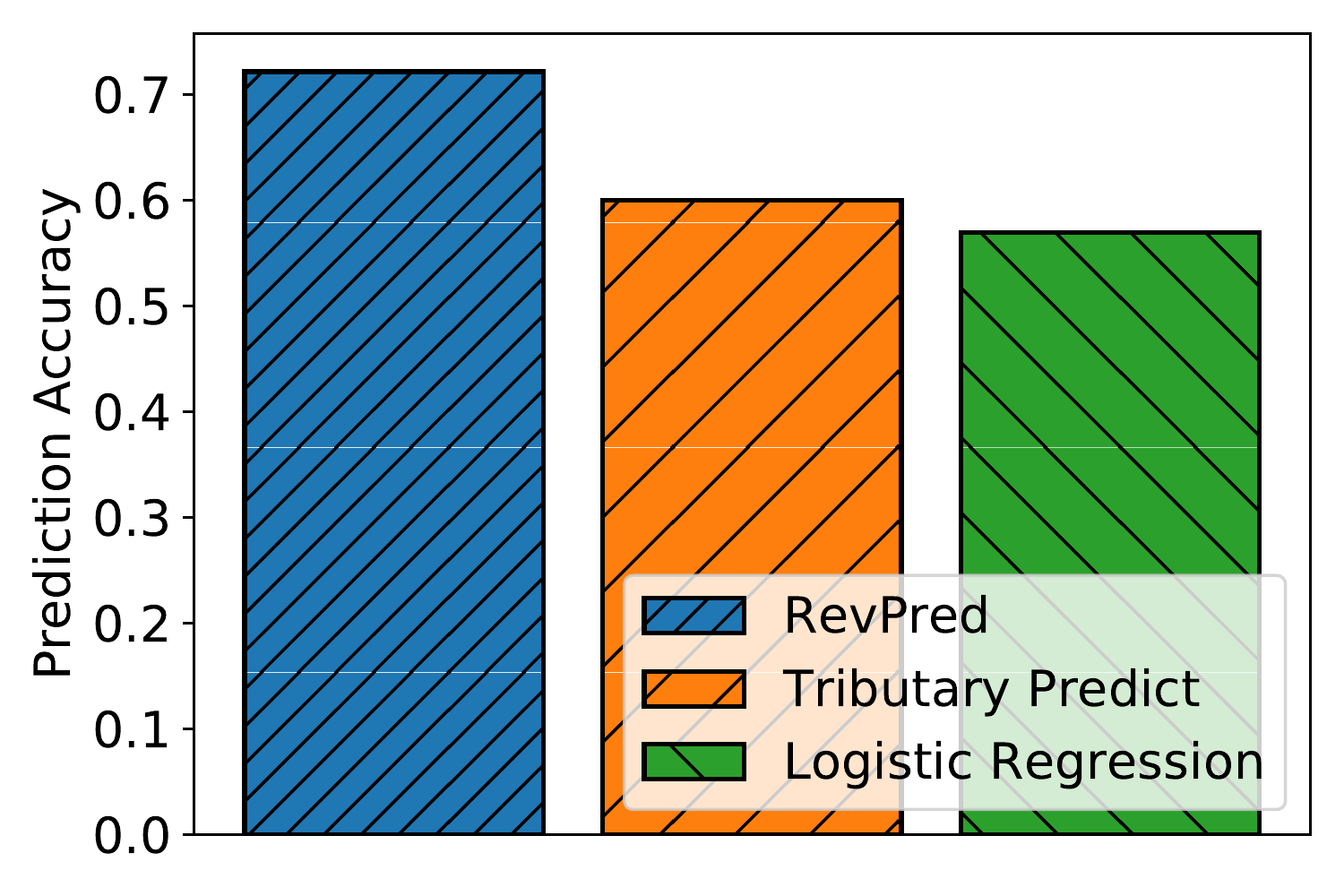}
		}
		\subfigure[F1 Score.]{
			\label{fig_pred_f1}
			\includegraphics[width=0.45\columnwidth]{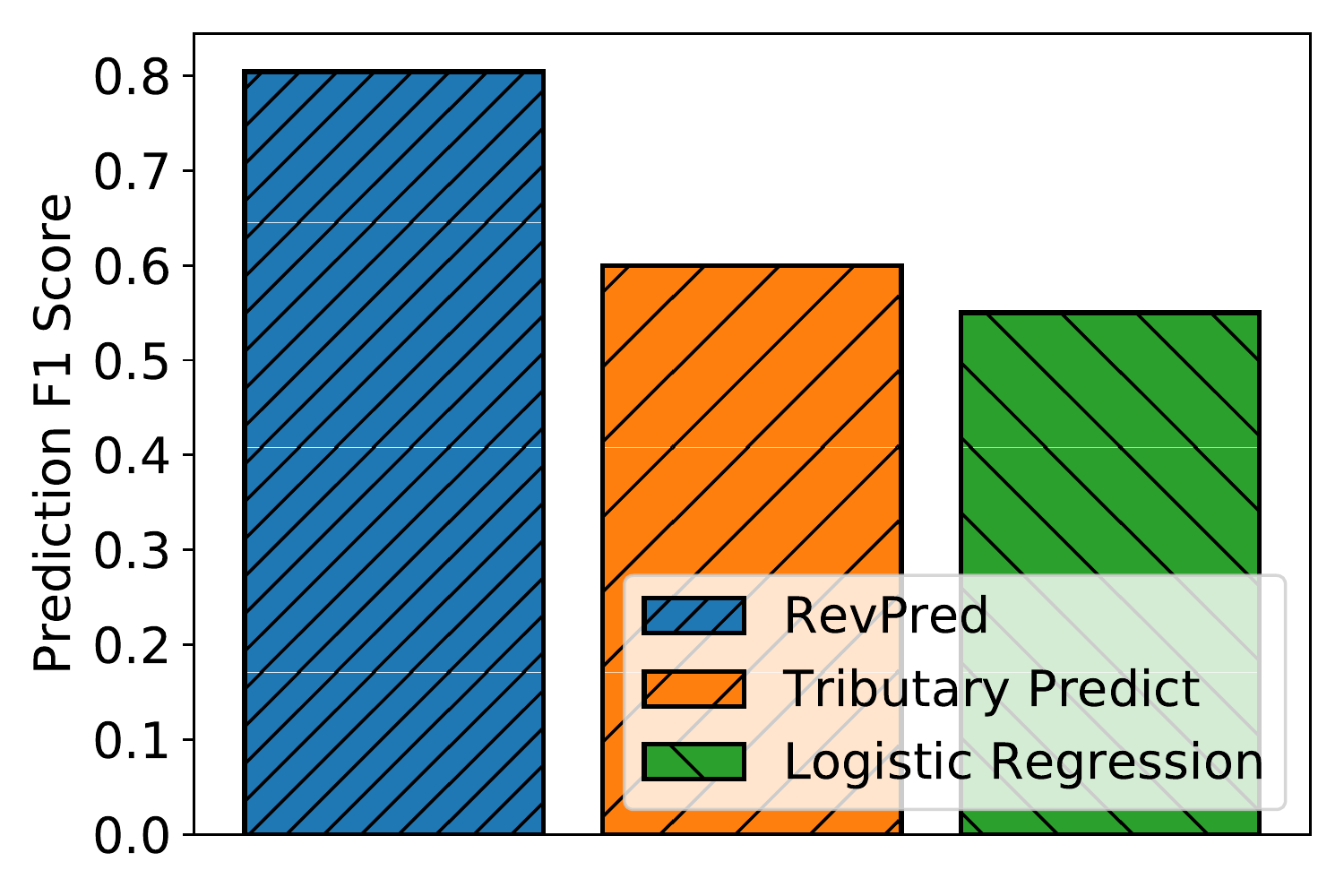}
		}
		\subfigure[Cost/PCR Comparison.]{
			\label{fig_pred_cmp}
			\includegraphics[width=.9\columnwidth]{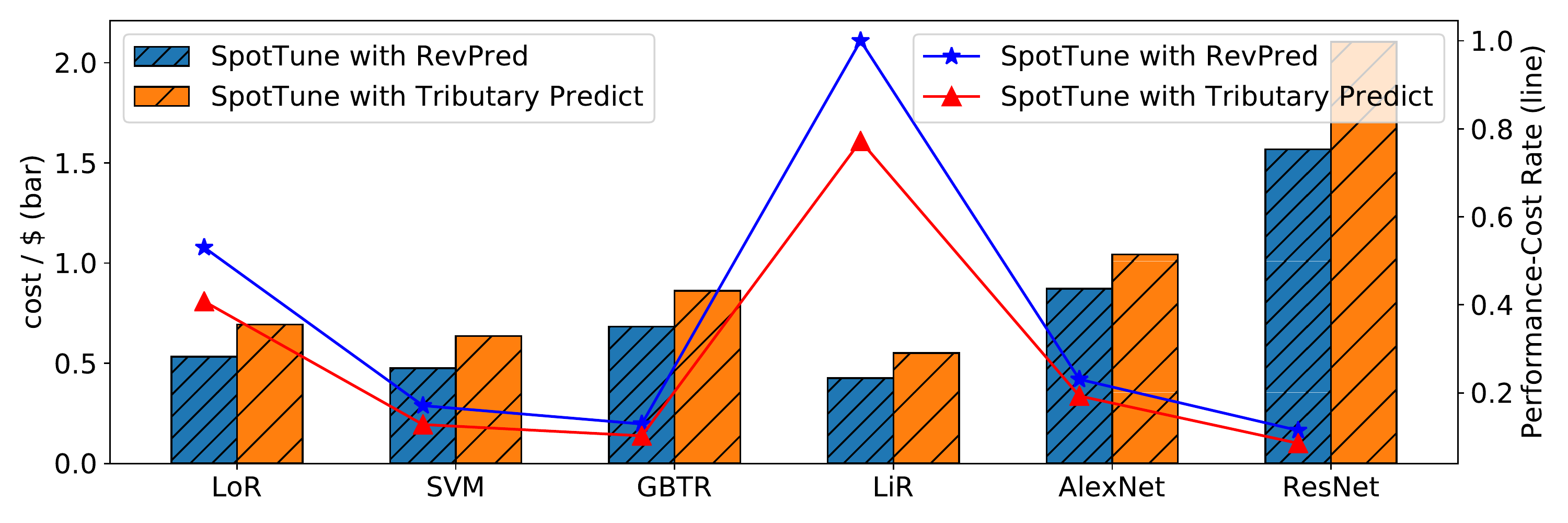}
		}
		\caption{Comparison between RevPred and Baselines}
		\label{fig_RevPred}
	\end{figure}

	This section evaluates the accuracy and effect of RevPred, which is described in Section \ref{RevPred_info}. The recent Tributary\cite{harlap2018tributary} system used the machine learning inference models trained with historical spot market data with engineered features and random-generated \textit{maximum price}. RevPred in SpotTune improves the prediction accuracy by \textbf{i)} separating the input data into history data and present data and \textbf{ii)} generating \textit{maximum price} according to the history price fluctuation when training the model, which we argue are the two reasons that RevPred outperforms Tributary. Since Tributary does not opensource its codes, we try to re-implement its prediction model from scratch, which is named as Tributary predict in our experiments. Figure \ref{fig_RevPred} shows the accuracy and F1 scores for three prediction models, RevPred, Tributary Predict, and Logistic Regression. These models are trained on spot market price data from 04/26/17-05/04/17 and are evaluated on data from 05/05/17-05/07/17 for instance types in Table \ref{tbl_conf}.
	
	The output of the prediction models is whether the instance specified with a price will be preempted within an hour. Accuracy scores are calculated by the \#samples predicted correctly divided by the total \#samples. F1 score is a synthetic accuracy measurement when the dataset is skewed. RevPred provides the best accuracy and the best F1 score. RevPred increases accuracy by 20.32\% and F1 score by 34.03\% compared with Tributary Predict.

	The increased accuracy and F1 scores are transferred to the effectiveness of SpotTune because the accurate predicting enables SpotTune to provision instances with lower \textit{step cost} more precisely. We integrate two spot instance revocation probability predictors, RevPred and the Tributary predict, with SpotTune to compare the cost and normalized PCR. As Figure \ref{fig_pred_cmp} shows, when using RevPred, SpotTune yields about 25\% less cost and 24\% more PCR compared to using the Tributary predictor.

	\subsection{Training Trend Predicting Effect}
	
	\begin{figure}[b]
		\centering
		\subfigure[Prediction Example.]{
			\label{fig_fit_emp}
			\includegraphics[width=0.45\columnwidth]{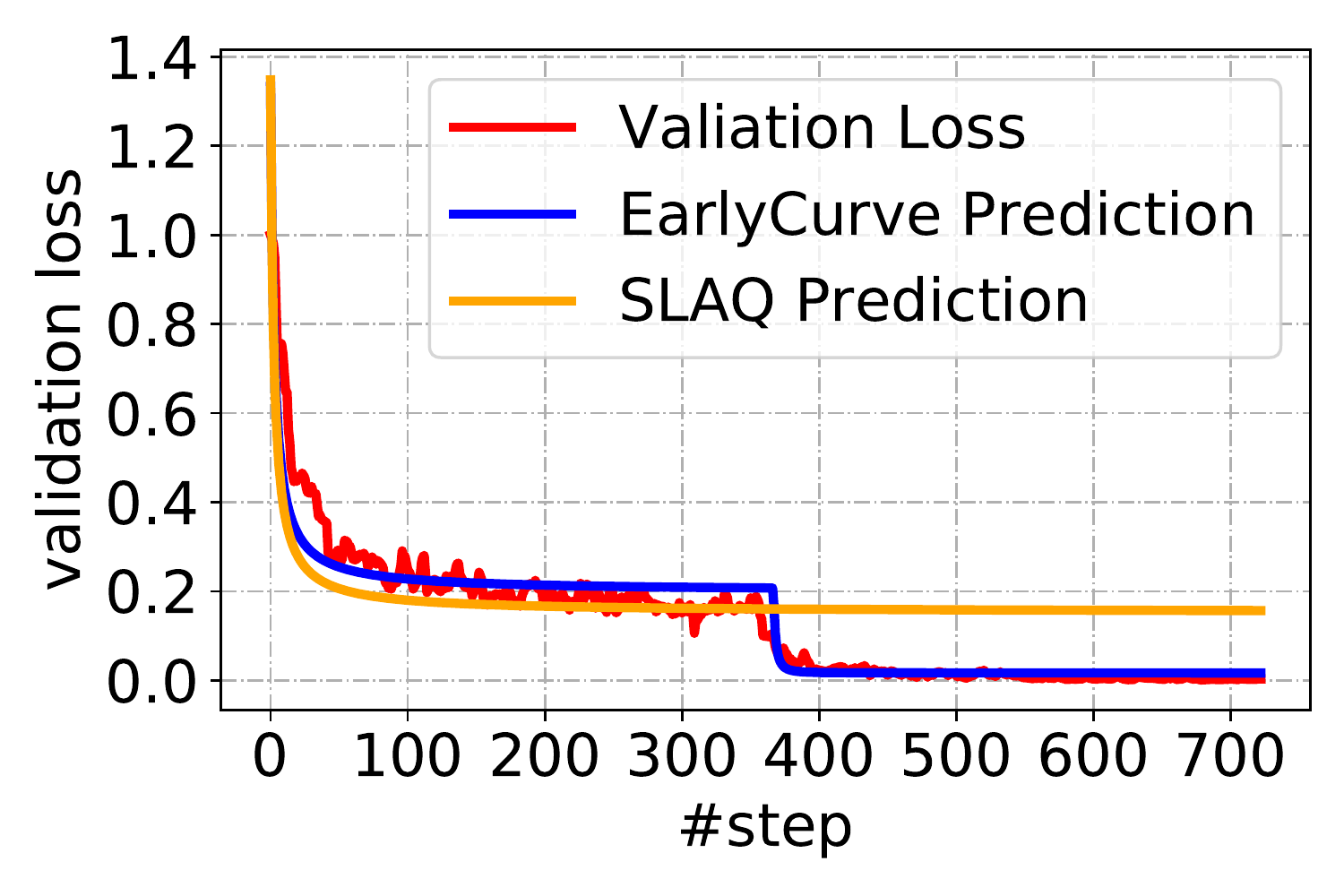}
		}
		\subfigure[Prediction Errors.]{
			\label{fig_fit_errors}
			\includegraphics[width=0.45\columnwidth]{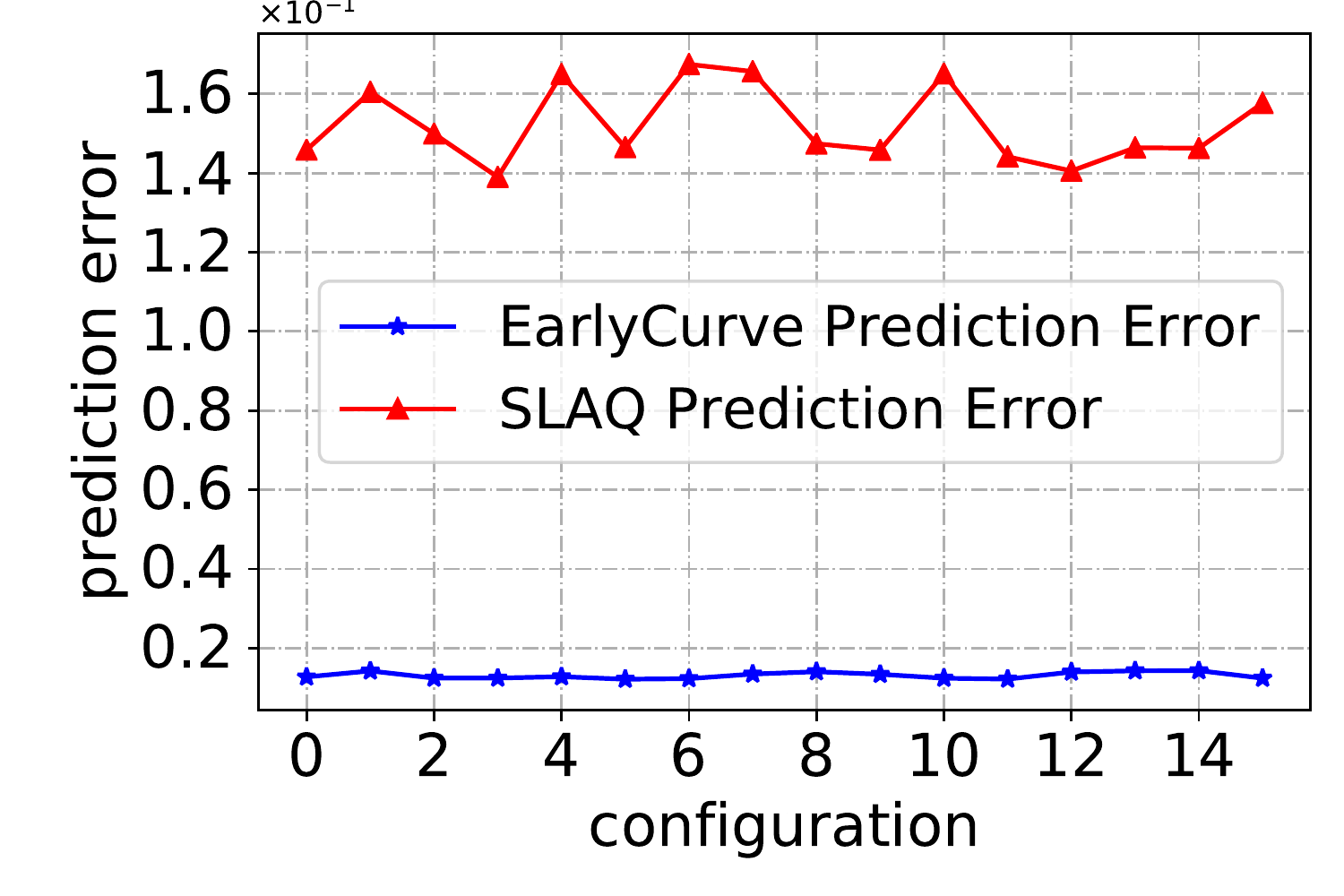}
		}
		\caption{Training Trend Prediction Effect.}
		\label{fig_fit}
	\end{figure}
	
	We compare EarlyCurve with previous training trending predicting method SLAQ\cite{zhang2017slaq} using the benchmark training metrics data with $\theta$ fixed as 0.7. Figure \ref{fig_fit_emp} shows an example of fitting curves using the two methods. As the baseline uses one-stage curve fitting, which does not take the multi-stage learning rate issue into consideration, its prediction error is significantly higher than EarlyCurve. Noted that if the learning rate of a model is not changing periodically, EarlyCurve and SLAQ would exhibit the same effect. Figure \ref{fig_fit_errors} shows the prediction error of the two methods on all 16 configurations of ResNet. The prediction accuracy of RevPred is shown in Figure \ref{fig_top3_acc_theta}.
	
	\subsection{Checkpoint-Restore Overhead}\label{sys_overhead}

	Checkpointing models to the remote object storage is the main overhead in SpotTune. We use the s3fs tool\cite{s3fs} to mount the remote object storage AWS S3 to a local directory on the machine so as to checkpoint the model conveniently. Besides, all the training data is restored on S3 before training as aforementioned, such that a new-started instance could instantly start to training without doing too much preparing work. Theoretically the maximum model size that could be uploaded to the remote storage would be the speed of checkpointing times 120 seconds because the notice is delivered two minutes before the real revocation as aforementioned.
	
	\begin{figure}[htbp]
		\centerline{\includegraphics[width=.45\textwidth]{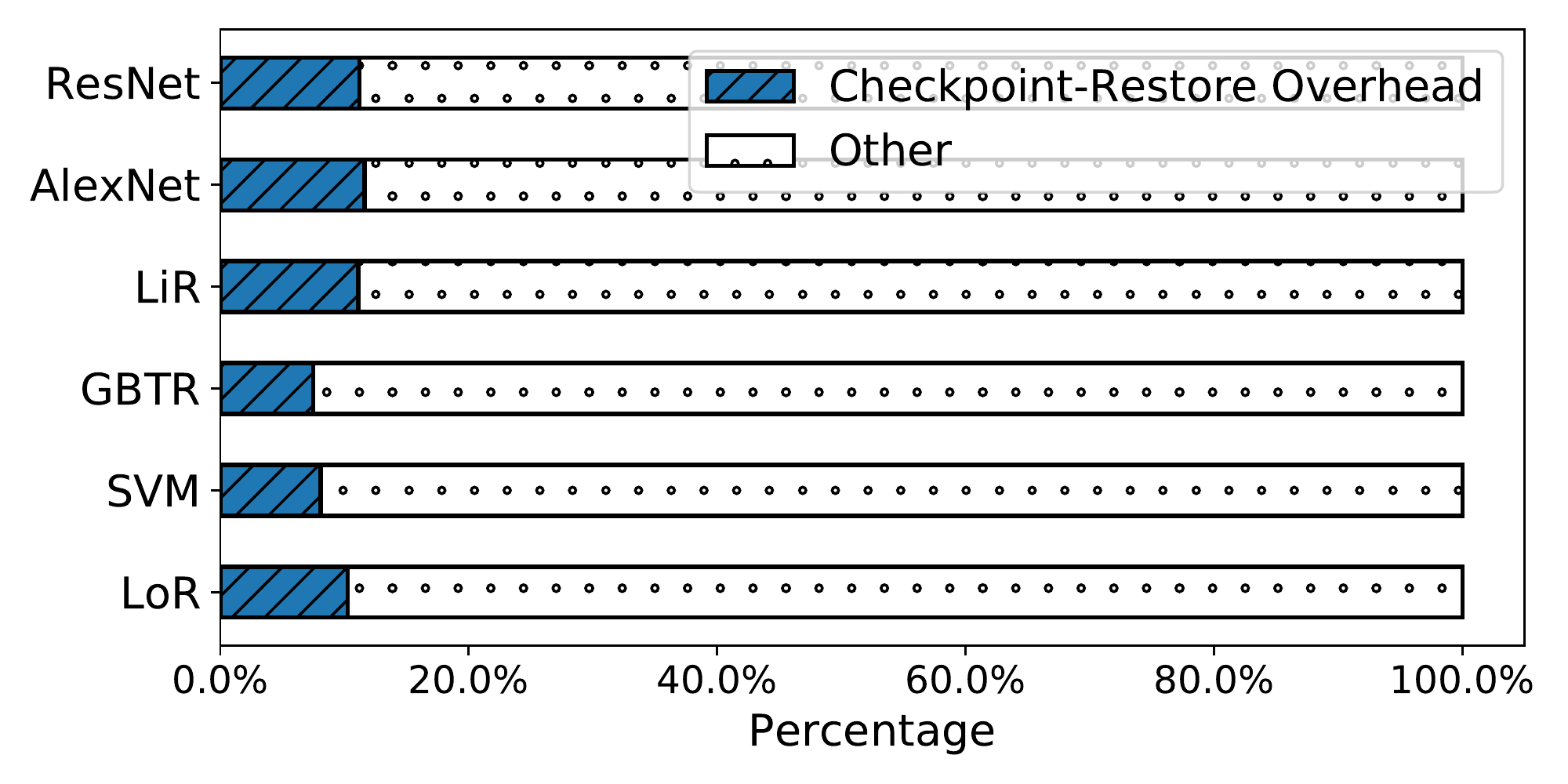}}
		\caption{Checkpoint-Restore Time Percentage.}
		\label{ckpt_overhead}
	\end{figure}
	

	We measure the speed of checkpointing models to AWS S3\cite{awss3} using Tensorflow on different types of instances. We observe that checkpointing is a CPU-bound process. The instance with the most cores, \textit{m4.4xlarge}, has the highest checkpointing speed 134.22MB/s and the largest max-model-size 15.73GB. For comparison we use the \textit{t2.micro} instance, with 1 CPU core and 1GB memory, as the testbed, achieving a 62.83MB/s speed and a 7.36GB model size upper bound. If the model size is over the maximum model size an instance supports, SpotTune could use periodically checkpointing or prediction-based checkpointing to handle it, which we leave for future work.
	
	Figure \ref{ckpt_overhead} shows the percentage of time fraction checkpoint-restore takes compared with the training time, less than 10\% on average. Compared with the training time and the cost-saving effect, we argue the checkpointing is worthy.
	
	\section{Discussion}\label{section_6}
	\subsection{Spot Instance Pool}\label{selecting_spot}
	According to Equation \ref{equation_step_cost} and our analysis in Subsection \ref{why_cheapest}, SpotTune may favor the unsteady spot markets such that could acquire more refund using our resource provisioning approach. But in the real world, spot markets exhibit various pricing patterns, among which some are very stable such that may not be preempted frequently. In this case, SpotTune degenerates to a trivial scenario without benefiting from the refund, i.e., the $p$ in Equation \ref{equation_step_cost} tends to zero. In this scenario, SpotTune would choose the instance with the lowest \textit{step cost}, without considering the revocation too much.
	
	\subsection{ML Model Convergence Rate}
	EarlyCurve in SpotTune models the ML training curve as a process converging at a speed of O($\frac{1}{k}$) or O($\frac{1}{k^2}$), namely sublinear. This magnitude of convergence rate is well-suited for gradient-descent-based ML optimization algorithms. But there are also some algorithms with linear or superlinear convergence rates, which converge at a rate of O($\mu^k$). For example, L-BFGS which is a widely used Quasi-Newton Method has a superlinear convergence rate which is between linear and quadratic\cite{zhang2017slaq}. Experimental ML benchmarks in this paper all use gradient descent as the optimization algorithms. 
	If the user uses SpotTune to select the best hyper-parameter setting of a superlinear converging model, a different curve-fitting model should be applied, which we will investigate in future work.
	
	\subsection{Non-converging ML Models}
	SpotTune's ML training trend prediction is based on the convergence property of the underlying optimizers with the metrics history data. Loss functions of some models initiated with bad hyper-parameters may not monotonically decrease, causing SpotTune to mispredict the training trend. In this case, the user may assign $\theta$ as 1 and indicate a proper metric representing the model's quality under a concrete hyper-parameter setting.
	
	\section{Related Work}\label{sec_related_work}
	\subsection{Hyper-Parameter Tuning}
	Many previous works\cite{luo2016review,lu2019hyper,shahriari2015taking,snoek2012practical,bergstra2012random,mantovani2015effectiveness,sparks2015automating} focus on improving the hyper-parameter search algorithms to reduce the times of attempting steps, while SpotTune looks at the underlying system/resources perspective to HPT. These algorithms could be integrated with SpotTune.
	
	\subsection{Leveraging Transient Resources}
	Many batch computing and elastic web service systems\cite{subramanya2015spoton,joaquim2019hourglass,shastri2017hotspot,harlap2018tributary,ali2019spotweb,harlap2017proteus} take advantage of transient resources to achieve a lower cost. SpotTune runs a new workload, HPT, on transient resources, which is different from them in the basic motivation. And SpotTune's resources acquisition mechanism is better than prior works such as Tributary\cite{harlap2018tributary}, which could also be applied to enhance the batch computing or elastic web service systems with modifications.

	Also, there are some works leveraging transient cloud resources for ML use at scale. Proteus \cite{harlap2017proteus} focuses on building a parameter server architecture using transient cloud resources for training an ML model in a distributed manner. SpotTune is different from it because it solves the problem of training models with different HP settings simultaneously rather than running a single model. MArk\cite{zhang2019mark} takes advantage of transient resources for cost benefits, but mainly focusing on the process of the model serving stage.
	
	\subsection{Spot Market Price Prediction}
	Prior works\cite{harlap2017proteus, harlap2018tributary, subramanya2015spoton} use different methods, including simple heuristic models and machine learning models, to predict the spot market pricing trend of AWS EC2. They either predict the future price of spot instances or the probability of revocation. The trend of the spot market price could be used to aid in the pro-active checkpointing or cost-efficient resources provisioning. RevPred in SpotTune improves the prediction accuracy by enhancing the data preprocessing approach and the model structure.
	
	\subsection{ML Training Prediction}
	Several systems predict the ML training process by either judging the model quality using a single-step metric\cite{li2016hyperband} or fitting the training curve using a single function\cite{zhang2017slaq, peng2018optimus}. They are suitable for limited scenarios because they miss the issue that the learning rate may have multiple stages. EarlyCurve in SpotTune uses staged function to handle this issue to reduce the prediction error.
	
	\section{Conclusion}\label{section_7}
	We present SpotTune, an orchestrating system leveraging transient resources in the public cloud for hyper-parameter tuning. SpotTune uses a fine-grained measurement \textit{step cost} to choose the cheapest instance to run a single step, thus reducing the overall cost and increase the performance-cost rate. SpotTune also applies an ML training trend predicting model to early shutdown the training process to save computation capacity. In the experiments, we prove that using SpotTune in the public cloud could make HPT both fast and cheap. SpotTune also has some limitations. In our future work,  we will consider enhancing SpotTune's support for checkpointing large models.
	
	\section*{Acknowledgments}
	This work is supported by the National Key R\&D Program of China (2016YFB1000105), the Science Fund for Creative Research Groups of China under Grant, No. 61421091. We also thank our shepherd, Purushottam Kulkarni, for help in shaping the final version of the paper.
	
	\bibliographystyle{IEEEtran}
	\bibliography{IEEEabrv,./paper}
	
\end{document}